\documentclass[11pt]{article}

\usepackage{graphicx}
\usepackage[top=1.25in,bottom=1.25in,left=1in,right=1in]{geometry}
\usepackage{amsmath,amsfonts,amscd,amssymb,bm,bbm,amssymb, prettyref} 
\usepackage[affil-it]{authblk}
\usepackage{amsthm}
\usepackage{algorithm, algorithmic}
\usepackage[usenames]{color}
\usepackage{mathtools}
\usepackage{tikz}
\usepackage{pgfplots}
\usepackage{dsfont}
\usepackage[toc,page]{appendix}
\usepackage{makecell}
\usepackage{caption} 
\usepackage{subcaption} 
\usepackage{booktabs}  
\usepackage{multirow}  
\usepackage{caption}
\usepackage{dsfont}

\usepackage{natbib} 
\usepackage{threeparttable}
\usepackage[subpreambles=true]{standalone} 
\usepackage{import} 
\definecolor{darkred}{RGB}{100,0,0}
\definecolor{darkgreen}{RGB}{0,100,0}
\definecolor{darkblue}{RGB}{0,0,150}
\definecolor{red}{RGB}{255,0,0}

\usepackage{hyperref}
\hypersetup{colorlinks=true, linkcolor=darkred, citecolor=darkgreen, urlcolor=darkblue}
\usepackage{url}

\newcommand{\Var}{\textrm{Var}}
\newcommand{\E}{\textrm{E}}
\newcommand{\Cov}{\textrm{Cov}}
\definecolor{xli}{RGB}{200,50,50}

\numberwithin{equation}{section}
\begin{document}
\title{Iteratively Reweighted Least Squares Method for Estimating Polyserial and Polychoric Correlation Coefficients}

\author[1]{Peng Zhang \thanks{Corresponding author.  E-mail: pengz@zju.edu.cn.}}%
\author[1]{Ben Liu}
\author[2]{Jingjing Pan}
\affil[1]{School of Mathematical Sciences, Zhejiang University,
Hangzhou, China, 310027.} 
\affil[2]{Zhejiang Super Soul Artificial Intelligence Research Institute}

\date{}
\maketitle
\begin{abstract}
An iteratively reweighted least squares (IRLS) method is proposed for estimating polyserial and polychoric correlation coefficients in this paper. It   iteratively calculates the slopes in a series of weighted linear regression models fitting on conditional expected values. For polyserial correlation coefficient,  conditional expectations of the latent predictor  is derived from the observed ordinal categorical variable, and the regression coefficient is obtained using weighted least squares method. In estimating polychoric correlation coefficient, conditional expectations of the response variable and the predictor are updated in turns.  Standard errors of the estimators are obtained using the delta method based on data summaries instead of the whole data. Conditional univariate normal distribution is exploited and a single integral is numerically evaluated in the proposed algorithm, comparing to the double integral computed numerically based on the bivariate normal distribution in the traditional maximum likelihood (ML) approaches. This renders the new algorithm very fast in estimating both polyserial and polychoric correlation coefficients. Thorough simulation studies are conducted to compare the performances of the proposed method with the classical ML methods. Real data analyses illustrate the advantage of the new method  in computation speed.
~\\

\textbf{keyword:} Iteratively reweighted least squares, Polyserial correlation, Polychoric correlation, Tetrachoric correlation, Maximum likelihood, Linear regression.
\end{abstract}

\section{Introduction}
In behavioural, educational and psychological studies, it is common that the observed variables are measured using ordinal scales. For example, Likert scale is widely used to measure responses in surveys, allowing individuals to express how much respondents  agree or disagree with a particular statement in a five (or seven) point scale. These categorical variables can be treated as being discretized from an underlying continuous variable for degree of agreement on the statement. There are also many examples of quantitative variables that are discretized explicitly in social science studies. For instance, when asking questions about sensitive or personal quantitative attributes (income, alcohol consumption), the non-response rate may often be reduced by simply asking the respondent to select one of two very broad categories(under \$30K/ over \$30K, etc.). When analyzing this kind of data, a common approach is to assign integer values to each category and proceed in the analysis as if the data had been measured on an interval scale with desired distributional properties.

The most common choice for the distribution of the latent variables is the normal distribution because all covariances between the latent variables can be fully captured by the covariance matrix and each of its elements can be estimated using a bivariate normal distribution separately. The correlation in the standard bivariate normal distribution is called tetrachoric correlation based on $2 \times 2$ contingency table was suggested by \cite{person1900polychoric}. The tetrachoric correlation was generalized to the case where the observed variables $X$ and $Y$ have $r$ and $s$ ordinal categories by \cite{ritchie1918correlation} and \cite{pearson1922polychoric} in the early 20th century, but it took over half a century before the computationally feasible maximum likelihood procedure was proposed by \cite{olsson1979maximum}. There have been two basic approaches to implementation: the so-called two-step method which first estimates the unknown thresholds from the marginal frequencies of the table and then finds the maximum likelihood estimate (MLE) of $\rho$ conditional on the estimated thresholds.  The second approach is to find the joint MLE of $(\rho, a, b)$ from the likelihood function. The author gives the equation system to be solved and, in addition, derives expressions for the information matrix which can be used to obtain asymptotic standard errors for the estimates.


Let $X$ be an observed ordinal variable which depends on an underlying latent continuous random variable $Z_1$ and $Y$ represent another observed continuous variable. It is assumed that the joint distribution of $Z_1$ and $Y$ is bivariate normal.  The product moment correlation between $X$ and $Y$ is called the point polyserial correlation, while the correlation between $Z_1$ and $Y$ is called the polyserial correlation. The MLE of the polyserial correlation has been derived by \cite{cox1974estimation}. 
\cite{olsson1982polyserial} derived the relationship between the polyserial and point polyserial correlation and compared the MLE of polyserial correlation with a two-step estimator and with a computationally convenient ad hoc estimator.

Another method to estimate tetrachoric and polychoric correlation coefficients is a Bayesian approach proposed by \cite{albert1992bayesian}. The author used a latent bivariate normal distribution to estimate a polychoric correlation coefficient from the Bayesian point of view by using the Gibbs sampler. One attractive feature of this method is that it can be generalized in a straightforward manner to handle a number of nonnormal latent distributions. They generalized their method to handle bivariate lognormal and bivariate $t$ latent distributions in their simulations.

\cite{chen2009comparison} and \cite{choi2011comparison} have showed that a different form of Bayesian estimation outperforms traditional maximum likelihood (ML) in a variety of settings, but their method is restricted only to the case of the bivariate Gaussian distribution. They correctly pointed out that, in real practice, the desirable sample sizes to obtain stable estimates for the polychoric correlation coefficient may not be available to the researcher. They claimed that due to the properties of numerical procedure of ML (i.e., iterative hill-climbing method using gradients of the target function), the ML estimation method for polychoric correlation coefficients has several disadvantages such as, local maxima, non-converged solution, an inaccurate estimation of the confidence interval and so on. Two new Bayesian estimates, maximum a posteriori (MAP) and expected a posteriori (EAP) are introduced and compared to ML method. In their simulation study, they found evidence that the MAP would be the estimator of choice for the polychoric correlation coefficients. 


Pearson correlations can be considered a less suitable method for studying the degree of association between categorical variables for several reasons. First, from a methodological point of view these variables would imply ordinal scales, whereas Pearson correlations assume interval measurement scales. Furthermore, the only information provided by this kind of scale is the number of subjects in each of the categories (cells) in a contingency table; if Pearson correlations are used in this case the relationship between measures would be artificially  restricted due to the restrictions imposed by categorization (\cite{gilley1993factor}), since all subjects situated in the interval that limits each of the categories would be considered as being included in the same category and, therefore, they would be assigned the same score with a resulting reduction in data variability.

\cite{holgado2010polychoric} illustrated the advantages of using polychoric rather than Pearson correlations in exploratory factor analysis(EFA) and confirmatory factor analysis(CFA), taking into account that the latter require quantitative variables measured in intervals, and that the relationship between these variables has to be monotonic. Their results showed that the solutions obtained by using polychoric correlations provide a more accurate reproduction of the measurement model used to generate the data.

More recently, network research has gained substantial attention in psychological sciences, which is called psychological networks by researchers. Psychological networks has been used in various different fields of psychology \cite{epskamp2018estimating}. The Gaussian graphical model(GGM) \cite{lauritzen1996graphical}, in which edges can directly be interpreted as partial correlation coefficients. The GGM requires an estimate of the covariance matrix as input, for which polyserial correlation and polychoric correlations can also be used in case the data are ordinal. However, for large network problems, it usually needs considerably longer computational time when using ML method.

In this paper we propose a simple and fast method to estimate the polyserial correlation coefficient and the  polychoric correlation coefficient. It is motivated by the fact that the Pearson's correlation coefficient coincides with the slope of the regression line for paired standard normal data. When one of the paired continuous data is discretized, an unbiased estimator of the slope is derived from the generated categorical data. When both of the paired data are discretized, the slope of the regression line, i.e. the correlation coefficient of the two normal random variables, will be obtained iteratively from a series of similar estimation procedures. The detail of the algorithm can be found in Section 2. In Section 3 and 4, we conduct simulation studies and data analyses to compare the proposed method with the ML method. At last, we conclude with discussions and some works can be done in the future  to improve the proposed method. 

\section{Iteratively Reweighted Least Squares Algorithm}

Assume  $(Z_1, Z_2)^T \sim N_2(\pmb 0, \pmb R)$ where $\pmb 0=(0,0)^{T}$ and 
$\pmb R=\left(
\begin{array}{cc}
1 & \rho \\
\rho & 1 
\end{array} 
\right)
$, $-1 \leq \rho \leq 1$. Conditioning on $Z_1$, $Z_2|Z_1 \sim N(\rho Z_1, 1- \rho^2)$. Hence 
\begin{equation}
\E(Z_2|Z_1) = \rho Z_1. \label{mod_simpreg}
\end{equation} 
This represents a simple linear regression model fitting $Z_2$ on $Z_1$ and $\rho$ is the slope  of the regression line.
Therefore, $\rho$, the Pearson correlation coefficient of $Z_1$ and $Z_2$,  can be estimated from such a linear regression model.

\subsection{Polyserial correlation coefficients} \label{sec:polyserial}
Consider the case where one of the paired random variables, namely $Z_1$, is discritized into an ordinal polychotomous variable, $X$, and the other is observed as a continuous variable, $Y$. Let $X$ be an observed ordinal variable with $s$ categories, generated from the latent variable $Z_1$ with $ X = i \text{~if~} a_{i-1} < Z_1 \leq a_i, i = 1, \dots, s,$ 
where $a_i$s are  thresholds with $a_{0} = -\infty$ and $a_{s} = \infty$. 

If $Z_1$ were observable, it would have been given from the regression line that $E(Y|Z_1) = \rho Z_1$. Taking expectation with respect to $Z_1$, 
\begin{equation}
\E\{\E(Y|Z_1)\} =  \rho \E(Z_1). \nonumber
\end{equation}
It holds for every $Z_1$ such that $a_{i-1} < Z_1 \leq a_i$, or correspondingly, $X=i$, for $i=1,2,\dots, s.$ That is, $$\E\{\E(Y|Z_1, a_{i-1} < Z_1 \leq a_i)\} =  \rho \E(Z_1|a_{i-1} < Z_1 \leq a_i),$$ or, 
\begin{equation}
\E\{\E(Y|X=i)\} =  \rho \E(Z_1|a_{i-1} < Z_1 \leq a_i),  \label{reg1}
\end{equation}
for $i = 1, \cdots, s$.

Denote $\E(Y|X=i)$ by $E_{Y_{i}}$ and $\E(Z_1|a_{i-1} < Z_1 \leq a_i)$ by $e_{x_i}$, equation (\ref{reg1}) is a regression model without an intercept, in which $E_{Y_{i}}$ is the response variable and  $e_{x_i}$ is the explanatory variable, with $\rho$ being the regression coefficient.  Because $E_{Y_{i}}$s have unequal variances, $\rho$ cannot be estimated with an ordinary least squares method. However, clearly $E_{Y_{i}}$s are independent to each other, $\rho$ can be estimated with a weighted least squares method with a diagonal weight matrix.

It is easy to show that the density function of $E_{Y_{i}}$ is
\begin{equation}
f(y) = \frac{\Phi\left(\frac{a_i - \rho y}{\sqrt{1-\rho^2}}\right)-\Phi\left(\frac{a_{i-1} - \rho y}{\sqrt{1-\rho^2}}\right)}{P_i}\phi(y),   \nonumber
\end{equation}
where $P_i = \textrm{Pr}(X = i)$. The mean and variance of $E_{Y_{i}}$, $\mu_i$ and $\sigma_i^2$, are given by
\begin{eqnarray}
    \mu_i &=& \rho\frac{\phi(a_{i-1})-\phi(a_i)}{P_i} \nonumber \\
    \sigma^2_i &=& 1 + \rho^2\frac{a_{i-1}\phi(a_{i-1})-a_i \phi(a_i)}{P_i} - \rho^2\frac{\{\phi(a_{i-1})-\phi(a_i)\}^2}{P_i^2} \label{sigmai}
\end{eqnarray}

Let   $y_{i_1}, y_{i_2}, \dots, y_{i_{n_i}}$  be the observed response variables associated with $X=i$  and $a_{i-1} < Z_{1j} \leq a_i$ for $j=1,\dots, n_i$, where $n_i$ is the size of data with $X=i$, $E_{Y_{i}}$ is estimated by 
\begin{equation}
\hat{E}_{y_i} = \bar{y}_{X=i} = \frac{1}{n_i}\sum_{j = 1}^{n_i} y_{i_j} \label{eyi}
\end{equation}
Since $Z_1$ has a truncated normal distribution with lower and upper limits $a_{i-1}$ and $a_i$ respectively, $e_{x_i}$ is the expected value of the truncated normal distribution,   given by
\begin{equation}
e_{x_{i}} = \frac{\phi(a_{i-1}) - \phi(a_i)}{P_{i}}, \label{defexi}
\end{equation}
where $\phi(\cdot)$ is the density function of the standard normal distribution. 
Let  $CP_{i}=\textrm{Pr}(X\leq i)=\Phi(a_i),~i = 1, \cdots, s$. Then 
\begin{eqnarray*}
CP_{i } = \sum_{j=1}^{i}{P_j} = \sum_{j=1}^{i}\{\Phi(a_{j}) - \Phi(a_{j-1})\},
\end{eqnarray*}
then $\hat{a}_i = \Phi^{-1}(\hat{CP}_{i})$, and $e_{x_i}$ in (\ref{defexi}) is estimated by 
\begin{equation}
\hat{e}_{x_{i}} = \frac{\phi(\hat{a}_{i-1}) - \phi(\hat{a_i})}{\hat{P}_{i}}  = \frac{\phi\{\Phi^{-1}(\hat{CP}_{i-1})\} - \phi\{\Phi^{-1}(\hat{CP}_{i})\}}{\hat{P}_{i}} \label{exi}
\end{equation}

Let $\mathbf{\hat{E}_x} = (\hat{e}_{x_1}, \hat{e}_{x_2}, \dots, \hat{e}_{x_s})^T$, $\mathbf{\hat{E}_y} = (\hat{E}_{y_1}, \hat{E}_{y_2}, \dots, \hat{E}_{y_s})^T$, and $$ \mathbf{\hat{\Sigma}} = 
\begin{bmatrix}
    \hat{\sigma}_1^2/n_1 & 0 & \ldots & 0\\
    0 & \hat{\sigma}_2^2/n_2 & \ldots & 0 \\
    \vdots & \vdots & \ddots & \vdots\\
    0 & 0 & \ldots & \hat{\sigma}_s^2/n_s
\end{bmatrix},
$$
the regression coefficient is given  by the weighted least squares method, 
\begin{equation}
\hat{\rho}=(\mathbf{\hat{E}_x}^T\mathbf{\hat{\Sigma}}^{-1}\mathbf{\hat{E}_x})^{-1}\mathbf{\hat{E}_x}^T\mathbf{\hat{\Sigma}}^{-1}\mathbf{\hat{E}_y}, \label{wls}
\end{equation}
 which is reduced to 
\begin{equation}
    \hat{\rho}=\frac{\sum_{i=1}^s n_i\hat{\sigma}_i^{-2}\hat{e}_{x_i}\hat{E}_{y_i}}{\sum_{i=1}^s n_i\hat{\sigma}_i^{-2}\hat{e}_{x_i}^2}.\label{rhoadj}
\end{equation}
While $\sigma^2_i$ in (\ref{sigmai}) depends on $\rho$, it can be obtained iteratively using the formula in (\ref{rhoadj}), with the Pearson correlation coefficient as the initial value.  The variance of $\hat{\rho}$ is given by
\begin{equation}
    \Var(\hat{\rho}) = (\mathbf{\hat{E}_x}^T\mathbf{\hat{\Sigma}}^{-1}\mathbf{\hat{E}_x})^{-1} = (\sum_{i=1}^sn_i\hat{\sigma}_i^{-2}\hat{e}_{x_i}^2)^{-1}, \label{varrho}
\end{equation}
and the standard error of $\hat{\rho}$ is $\sqrt{ \Var(\hat{\rho})}$.

The details of the IRLS  algorithm for estimating polyserial correlation coefficient are given in the following Algorithm \ref{polyserial},
\begin{algorithm}[H]\label{polyserial}
	\renewcommand{\algorithmicrequire}{\textbf{Input:}}
	\renewcommand{\algorithmicensure}{\textbf{Output:}}
	\caption{IRLS  algorithm to compute polyserial correlation}
	\label{polyserial}
	\begin{algorithmic}[1]
		\REQUIRE observed continuous variable $y$ and  ordinal variable $x$
		\ENSURE $\hat{\rho}$, polyserial correlation coefficient of $Y$ and $X$; $\sqrt{\Var\hat{\rho}}$, the standard error of $\hat{\rho}$
		\STATE Compute $\hat{P}_i,\hat{CP}_{i}$from data $x$.Estimate the thresholds $\hat{a}_i = \Phi^{-1}(\hat{CP}_{i}), i = 1, \dots, s$
		\FOR{$i=1$ to $s$} 
		\STATE Compute $\hat{e}_{x_{i}}$ using the  formula in equation (\ref{exi}) \\ 
		\STATE Compute $\hat{E}_{y_i}$ using the formula in equation (\ref{eyi})
		
		\ENDFOR 
		\STATE Compute the initial $\hat{\rho}$ using the Pearson correlation coefficient between $y$ and $x$ 
		\STATE Set $\text{iter} = 0$, $\text{diff} = 1$, $n = 100$, $\epsilon = 1e{-8}$ and $\hat{\rho}_0=\hat{\rho}$
        \WHILE{$\text{iter} < n ~\&~  \text{diff} > \epsilon$}

        \FOR{$i=1$ to $s$} 
        \STATE Compute $\hat{\sigma}^2_i$ using the formula in equation (\ref{sigmai})with $\hat{\rho}_0$
        \ENDFOR
        \STATE Update $\hat{\rho}$ using the formula in equation(\ref{rhoadj})
        \STATE Compute $\text{diff} = \hat{\rho} - \hat{\rho}_0$
        \STATE Update $\hat{\rho}_0 = \hat{\rho}$ and iter = iter + 1
        \ENDWHILE
        \STATE Compute $\Var(\hat{\rho})$ using the formula in equation(\ref{varrho})
        \STATE \textbf{return} $\hat{\rho}$ and $\sqrt{\Var\hat{\rho}}$
	\end{algorithmic}  
\end{algorithm}

\subsection{Tetrachoric  and polychoric correlation coefficients}
In this section, we extend the weighted least squares method for estimating polyserial correlation coefficients to an iteratively reweighted least squares method for estimating tetrachoric and polychoric correlation coefficients. When both of the paired normal variables are discritized into ordinal variables, neither the response nor the predictor variable is observable. Both the weights and the response parts in the formula (\ref{wls}) have to be updated after $\hat{\rho}$ is obtained in each iteration. Hence we call it the iteratively reweighted least squares algorithm. The correlation coefficient between observed categorical variables is smaller than that between the latent continuous variables in magnitude. However, the same procedure given in Section \ref{sec:polyserial} will update the estimate of the correlation coefficient, making it closer to the true parameter than the previous one. Therefore, the polychoric correlation coefficient can be estimated iteratively by updating conditional expectations of the two latent variables in turns. The series of the estimates will converge to the polychoric correlation coefficient. 
 
 Let $(Z_1, Z_2)^{T} \sim N_2(\pmb 0, \pmb R)$ where $\pmb 0=(0,0)^{T}$ and 
$\pmb R=\left(
\begin{array}{cc}
1 & \rho \\
\rho & 1 
\end{array} 
\right)
$, $-1 \leq \rho \leq 1$. $Z_1$ and $Z_2$ are categorized into $X$ and $Y$ with $s$ and $r$ categories respectively. 
After the categorization,  only $X$ and $Y$, instead of $Z_1$ and $Z_2$, are observable. We aim to estimate the correlation coefficient between $Z_1$ and $Z_2$ by the observed data $X$ and $Y$, which is called polychoric correlation. It is also known as tetrachoric correlation coefficient when $s=r=2$. Assume that  $X$ is generated from the latent variable $Z_1$ with $ X = i \text{~if~} a_{i-1} < Z_1 \leq a_i, i = 1, \dots, s,$ where $a_i$s are  thresholds with $a_{0} = -\infty$ and $a_{s} = \infty$. Similarly, $Y$  is generated from $Z_2$ with $ Y = j \text{~if~} b_{j-1} < Z_2 \leq b_j, j = 1, \dots, r,$ where $b_j$ are  thresholds with $b_{0} = -\infty$ and $b_{r} = \infty$. 
The data usually consist of an array of observed frequencies $n_{ij}: i = 1, \dots, s; j = 1, \dots, r$, which is the number of $X = i, Y = j$. Let $P_{ij} = \text{P}(X = i, Y = j) = \text{P}(a_{i-1} < Z_1 \leq a_i, b_{j-1} < Z_2 \leq b_j)$ be the proportion of data in  cell $(i, j)$. The cumulative marginal proportions of the table are $P_{i\cdot} = \sum_{k=1}^{i}\sum_{j=1}^{r} P_{kj}, i = 1, \dots, s$, $P_{\cdot j} = \sum_{i=1}^{s} \sum_{k=1}^{j}P_{ik}, j = 1, \dots, r$.

The IRLS method for estimating the tetrachoric correlation coefficient is derived first. Suppose that $X$ and $Y$ are two dichotomous variables,  generated from $Z_1$ and $Z_2$ respectively. Let $\mathbf{N} = \left(\begin{array}{cc} N_{11} & N_{12} \\ N_{21} & N_{22} \end{array} \right)$ be the contingency table with $N_{ij}, ~ i , j  = 1, 2 $ the frequencies of $X$ by $Y$ at the category $i$ and $j$ respectively, $N = N_{11} + N_{21} + N_{12} + N_{22}$. 

Since $Z_2$ is not observable, the regression model in (\ref{reg1}) becomes
\begin{equation}
\E\{\E(Z_2|X=i)\} =  \rho \E(Z_1|a_{i-1} < Z_1 \leq a_i),  ~ i = 1, 2.  \label{reg2}
\end{equation}

Denote the explanatory variables on the right hand side of (\ref{reg2}) by $e_{x_{i}}$. It is given by
\begin{equation} 
\begin{split}
e_{x_{1}} &\hat{=} \E(Z_1|Z_1 \leq a) = -\frac{\phi(a)}{P_{1\cdot}} = -\frac{\exp(-\frac{a^2}{2})}{\sqrt{2\pi}P_{1\cdot}},\label{ab} \\
e_{x_{2}} &\hat{=} \E(Z_1|Z_1 > a) = \frac{\phi(a)}{1-P_{1\cdot}} = \frac{\exp(-\frac{a^2}{2})}{\sqrt{2\pi}(1-P_{1\cdot})}.
\end{split}
\end{equation}
where $a = \Phi^{-1}(P_{1 \cdot})$ and $b = \Phi^{-1}(P_{\cdot 1})$ are the cutoff points, $\hat{P}_{1 \cdot} = \frac{N_{11} + N_{12}}{N}$ is the proportion of $X = 1$ and $\hat{P}_{\cdot 1} = \frac{N_{11} + N_{21}}{N}$ is the proportion of $Y = 1$. Similarly, $\hat{P}_{2 \cdot}$ and $\hat{P}_{\cdot 2} $ are proportions of $X = 2$ and $Y = 2$ respectively. $e_{x_{1}}$ and $e_{x_{2}}$ are estimated with formulae in equations (\ref{ab}) by plugging in the observed frequencies of the contingency table, and are used as the  initial value of predictors in the regression model (\ref{reg2}). 

Denote the response variable of the regression model in (\ref{reg2}) by $E_{Y_{i}} = \E(Z_2|X = i),~ i = 1, 2$, or equivalently, $\E(Z_2|Z_1 \leq a)$ and $\E(Z_2|Z_1 > a)$. It cannot be calculated directly. But because $Z_2$  is dichotomized by whether $Z_2 \leq b$ or  $Z_2 > b$ into two categories, indicated by either $Y =1$ or $Y =2$, $ E_{Y_{i}}$ can be  obtained based on the binomial distribution of $Y$, or into which of the two intervals $Z_2$ falls. This procedure is presented in the following two steps:

Firstly,  $Z_1$ and $Z_2$ are jointly normally distributed, $Z_2|Z_1 \sim N(\rho Z_1, 1-\rho^2)$. Standardization gives $\tilde{Z}_2=\frac{Z_2-\rho Z_1}{\sqrt{1-\rho^2}} \sim N(0,1)$. Denote $ \frac{b-\rho Z_1}{\sqrt{1-\rho^2}} $ by $Z_1^{b}$, then 
\begin{eqnarray*}
\E(\tilde{Z}_2 | Z_2 \leq b) & = & \E\left(\tilde{Z}_2 | \tilde{Z}_2 \leq Z_1^{b}\right) = -\phi\left(Z_1^{b}\right)/\Phi \left(Z_1^{b}\right), \\
\E(\tilde{Z}_2 | Z_2 > b) & = & \E\left(\tilde{Z}_2 | \tilde{Z}_2 > Z_1^{b} \right) = \phi\left(Z_1^{b}\right)/\left\{1-\Phi \left(Z_1^{b}\right)\right\}.  
\end{eqnarray*}

Inverting the standardization gives 
\begin{eqnarray*}
\E(Z_2|Z_1, Z_2 \leq b) & = &  \rho Z_1 - \sqrt{1-\rho^2} \phi \left(Z_1^{b} \right)/\Phi \left(Z_1^{b} \right),  \\ 
\E(Z_2|Z_1, Z_2 > b) & = & \rho Z_1 + \sqrt{1-\rho^2} \phi  \left(Z_1^{b} \right) /\left\{1-\Phi (Z_1^{b}) \right\}.
\end{eqnarray*}

Then the conditional mean of $Z_2$ given $Z_1$ at $e_{x_{1}}$ and $e_{x_{2}}$ for different categories of $Y$  are  
\begin{eqnarray} \nonumber
e_{11} &=& \E(Z_2|Z_2 \leq b, Z_1 = e_{x_{1}}) = \rho e_{x_1} - \sqrt{1-\rho^2} \phi \left(e_{x_1}^{b}\right)/\Phi \left(e_{x_1}^{b}\right),\\  \label{eyx}
e_{21} &=& \E(Z_2|Z_2 \leq b, Z_1 = e_{x_{2}}) = \rho e_{x_2} - \sqrt{1-\rho^2} \phi \left( e_{x_2} ^{b}\right)/\Phi \left( e_{x_2} ^{b}\right),\\ \nonumber
e_{12} &=& \E(Z_2|Z_2 > b, Z_1 = e_{x_{1}}) = \rho e_{x_1} + \sqrt{1-\rho^2} \phi \left(e_{x_1}^{b}\right)/\left\{1-\Phi \left(e_{x_1}^{b}\right) \right\},\\ \nonumber
e_{22} &=& \E(Z_2|Z_2 > b, Z_1 = e_{x_{2}}) = \rho e_{x_2} + \sqrt{1-\rho^2} \phi \left( e_{x_2} ^{b}\right)/\left\{1-\Phi \left( e_{x_2} ^{b}\right) \right\}, 
\end{eqnarray}
where $\phi(\cdot)$ and $\Phi(\cdot)$ are the density function and the cumulative distribution function of the standard normal distribution  respectively. 

Secondly, $E_{Y_{i}}$ is obtained based on conditional expected values in (\ref{eyx}). Because $Z_2$ is dichotomized into $Y$,  with two mass probabilities, then
\begin{equation}\label{ey}
\begin{split}
E_{Y_1} = \frac{P_{11}}{P_{11} + P_{12}}e_{11} +  \frac{P_{12}}{P_{11} + P_{12}}e_{12}, \\
E_{Y_2} = \frac{P_{21}}{P_{21}+P_{22}}e_{21} +  \frac{P_{22}}{P_{21}+P_{22}}e_{22}.
\end{split}
\end{equation}
where $P_{ij}, i=1,2; j=1,2$ is the proportion in cell $(i, j)$ of the probability table: 
$$\mathbf{P} = \left(\begin{array}{cc} P_{11} & P_{12} \\ P_{21} & P_{22} \end{array} \right).
$$

Similar as  in model (\ref{reg1}), $E_{Y_{i}}$s have unequal variances. They are not independent, yet the regression coefficient can be obtained with a weighted least square method, as long as the covariance matrix of the response variables is available. The variance  of $\hat{P}_{ij}$, the observed  proportion  of data in cell $(i, j)$, is given by     
\begin{equation}\label{var1}
    \Var(\hat{P}_{ij}) = \frac{1}{N^2}\sum_{k=1}^N \Var \{\mathbb{I}(X_k=i,Y_k=j)\} = \frac{1}{N}P_{ij}(1-P_{ij}), 
 \end{equation}
 and the covariance between $\hat{P}_{i_1j_1}$ and $\hat{P}_{i_2j_2}$ is given by
 \begin{eqnarray} 
    \Cov(\hat{P}_{i_1j_1},\hat{P}_{i_2j_2}) 
    &=& \frac{1}{N^2} \Cov\left\{\sum_{k=1}^N \mathbb{I}(X_k=i_1,Y_k=j_1),\sum_{k=1}^N \mathbb{I}(X_k=i_2,Y_k=j_2)\right\} \nonumber \\
    &=& \frac{1}{N^2}\{(N^2-N)P_{i_1j_1}P_{i_2j_2}-N^2P_{i_1j_1}P_{i_2j_2}\} \nonumber \\
    &=& -\frac{1}{N}P_{i_1j_1}P_{i_2j_2},~(i_1,j_1)\neq(i_2,j_2). \label{cov}
\end{eqnarray}

Denote the covariance matrix of  $\hat{\mathbf{P}}$ by $\mathbf{B}$.Then 
\begin{equation}\label{cov1}
\mathbf{B} =
    \left(
        \begin{array}{cccc} 
        P_{11}(1-P_{11}) & -P_{11}P_{12} & -P_{11}P_{21} & -P_{11}P_{22}\\  
        -P_{12}P_{11} & P_{12}(1-P_{12}) & -P_{12}P_{21} & -P_{12}P_{22}\\  
        -P_{21}P_{11} & -P_{21}P_{12} & P_{21}(1-P_{21})  & -P_{21}P_{22}\\
        -P_{22}P_{11} & -P_{22}P_{12} & -P_{22}P_{21} & P_{22}(1-P_{22})
        \end{array} 
    \right)/N.
\end{equation}

$E_{Y_{i}}$ is multivariate differentiable function of ${\mathbf{P}}$. The partial derivative matrix of $E_{Y_{i}}$ with respect to $\mathbf{P}$, $  \mathbf{D} $,  is listed in the appendix. 

The covariance matrix of vector $\mathbf{E_{Y}} = (E_{Y_{1}}, E_{Y_{2}})^T$, is given by $\mathbf{\Sigma} =  \mathbf{D}  \mathbf{B} \mathbf{D}^ \prime$, using the delta method. $\hat{\mathbf{\Sigma}}$ can be obtained  by replacing all $P_{ij}$ with the observed frequencies $\hat{P}_{ij}$. Let $\mathbf{e_x} = (e_{x_1}, e_{x_2})^T$, the regression coefficient in model (\ref{reg2}) can be calculated using a weight least square method as:

\begin{equation} 
        \hat{\rho} = (\mathbf{\hat{e}_x}'\hat{\mathbf{\Sigma}}^{-1}\mathbf{\hat{e}_x})^{-1}\mathbf{\hat{e}_x}'\hat{\mathbf{\Sigma}}^{-1}\mathbf{\hat{E}_{Y}}, \label{rho2}
\end{equation}
and 
\begin{equation} 
        \Var (\hat{\rho} ) = (\mathbf{\hat{e}_x}'\hat{\mathbf{\Sigma}}^{-1}\mathbf{\hat{e}_x})^{-1}, \label{varrho2}
\end{equation}
The standard error of $\hat{\rho}$ is $\sqrt{\Var(\hat{\rho})}$.


The predictors in equation (\ref{reg2}) are not constants. They have to be updated with the improved estimate $\hat{\rho}$ from (\ref{rho2}). This also consists of the following two steps. First, obtain the conditional expected values of $Z_1$ given $Z_2$ for different categories of $Z_1$, using a similar derivation of (\ref{eyx} -- \ref{ey}),

\begin{equation}\label{ey1}
    \begin{split}
    E_{Y_{x1}} = \frac{P_{11}}{P_{11} + P_{21}}e_{11} +  \frac{P_{21}}{P_{11} + P_{21}}e_{21}, \\
    E_{Y_{x2}} = \frac{P_{12}}{P_{12}+P_{22}}e_{12} +  \frac{P_{22}}{P_{12}+P_{22}}e_{22},
    \end{split}
\end{equation}

\begin{equation} \label{exy}
    \begin{split}
    e_{x11} &\hat{=} \E(Z_1|Z_1 \leq a, Z_2 = E_{Y_{x1}} ) = \rho E_{Y_{x1}}  - \sqrt{1-\rho^2} \phi \left(E_{Y_{x1}}^a \right)/\Phi \left(E_{Y_{x1}}^a \right), \\
    e_{x12} &\hat{=} \E(Z_1|Z_1 \leq a, Z_2 = E_{Y_{x2}} ) = \rho E_{Y_{x2}} - \sqrt{1-\rho^2} \phi \left(E_{Y_{x2}}^a \right)/\Phi \left(E_{Y_{x2}}^a \right),\\
    e_{x21} &\hat{=} \E(Z_1|Z_1 > a, Z_2 = E_{Y_{x1}}) = \rho E_{Y_{x1}} + \sqrt{1-\rho^2} \phi \left(E_{Y_{x1}}^a \right)/\left\{1-\Phi \left(E_{Y_{x1}}^a \right) \right\},\\
    e_{x22} &\hat{=} \E(Z_1|Z_1 > a, Z_2 = E_{Y_{x2}}) = \rho E_{Y_{x2}} + \sqrt{1-\rho^2} \phi \left(E_{Y_{x2}}^a \right)/ \left\{1-\Phi \left(E_{Y_{x2}}^a \right) \right\},
    \end{split}
\end{equation}
Second, update $e_{x_1}$ and $e_{x_2}$ by taking expectation over $Z_2$,
\begin{equation}\label{ex}
\begin{split}
e_{x_{1}} & = \frac{P_{11}}{P_{11}+P_{12}}e_{x11} +  \frac{P_{12}}{P_{11}+P_{12}}e_{x12},\\
e_{x_{2}} & = \frac{P_{21}}{P_{21}+P_{22}}e_{x21} +  \frac{P_{22}}{P_{21}+P_{22}}e_{x22}.
\end{split}
\end{equation}

\begin{figure}[H]
\centering
\includegraphics[width=0.6\textwidth,clip]{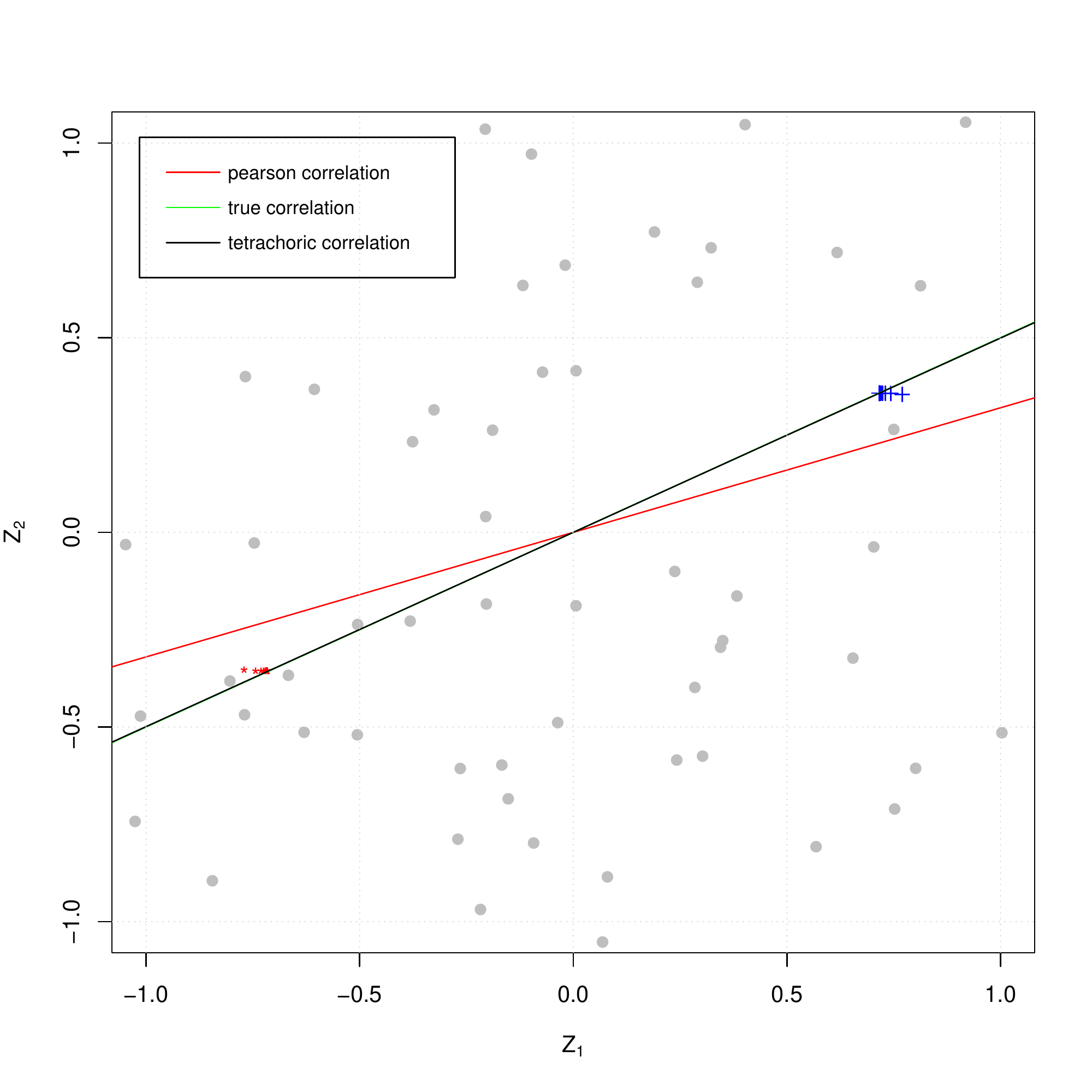}
\caption{The procedure of obtaining the tetrachoric correlation coefficient of IRLS  algorithm}\label{vertaxc}
\end{figure}

Then we go back to (\ref{eyx}) and repeat the previous procedure until $\hat{\rho}$ converges. Because both the weight matrix  and the response vector in equation (\ref{rho2}) have been updated, we call this an iteratively reweighted least squares algorithm. The proposed IRLS algorithm proceeds in sequence: $\mathbf{e_x} \rightarrow \mathbf{E_{Y}}, \mathbf{\Sigma} \rightarrow \rho \rightarrow \mathbf{e_x} \cdots \rightarrow \mathbf{E_{Y}},  \mathbf{\Sigma} \rightarrow \rho , \Var( \rho)$. Figure \ref{vertaxc} shows the change of the values used in (\ref{rho2}). 

The red star points are the $(\hat{e}_{x_{1}}, \hat{E}_{ Y_{1}}) $ and the blue points are $(\hat{e}_{x_{2}},  \hat{E}_{ Y_{2}}) $. It can be seen that  the algorithm converges fast and stops in several iterations. The slope of the red line is the Pearson correlation of the observed categorical data. It is apparently  smaller than the true $\rho$. We use it as the initial value of in the algorithm. The slope of the green line is the true value of latent variable $Z_1$ and $Z_2$. The slope of the black line is the estimated tetrachoric correlation coefficient from the proposed IRLS  algorithm. Clearly, as the algorithm proceeding, the results of our algorithm is getting closer to the true value. The details of the IRLS  algorithm for estimating tetrachoric correlation coefficient are given in the following Algorithm \ref{tetra}, 

\begin{algorithm}[H] 
	\renewcommand{\algorithmicrequire}{\textbf{Input:}}
	\renewcommand{\algorithmicensure}{\textbf{Output:}}
	\caption{IRLS method to compute tetrachoric correlation coefficient}
	\label{tetra}
	\begin{algorithmic}[1]
		\REQUIRE observed ordinal data $y$ and $x$ or contingency table
		\ENSURE tetrachoric correlation coefficient $\hat{\rho}$ of $Y$ and $X$, $\mbox{s.e.}(\hat{\rho})$ 
        \STATE Calculate the frequency table $\hat{\mathbf{P}}=(\hat{P}_{ij}),i,j=1,2$, the covariance $\mathbf{\hat{B}}$ using equation (\ref{cov1})
		\STATE Estimate the thresholds given from the cumulative marginal proportions of $\hat{\mathbf{P}}$.
		$\hat{a}_i = \Phi^{-1}(\hat{P}_{\cdot 1})$,
		$\hat{b}_j = \Phi^{-1}(\hat{P}_{1\cdot })$.
		\STATE Initialization: \\
		$\hat{\rho}_0$ = Pearson correlation coefficient of $x$ and $y$,\\
		$\hat{e}_{x_{1}}$ and 
		$\hat{e}_{x_{2}}$ using the formulae in equations (\ref{ab})
		\STATE Set $\text{iter} = 0$, $\text{diff} = 1$, $n = 100$ and $\epsilon = 1e{-8}$
		\WHILE{$\text{iter} < n ~\&~  \text{diff} > \epsilon$}
		\STATE Compute $\hat{E}_{y_{1} }$, $\hat{E}_{y_{2}} $  using equations~(\ref{eyx}) and (\ref{ey})
        \STATE Compute the derivation matrix $\mathbf{\hat{D}}$ using equation~(\ref{dera}), $\mathbf{\hat{\Sigma} }= \mathbf{\hat{D}}\mathbf{\hat{B}}\mathbf{\hat{D}'}$
		\STATE Compute $\hat{\rho}$ using the formula in equation (\ref{rho2}) 
		\STATE Compute $\text{diff} = \hat{\rho} - \hat{\rho}_0$, 
		\STATE Update $\hat{e}_{x_1}$ and $\hat{e}_{x_2}$ using the formula in equations~(\ref{ey1})(\ref{exy}) and (\ref{ex})
		\STATE Update $\hat{\rho}_0 = \hat{\rho}$ and iter = iter + 1
		\ENDWHILE
        \STATE Compute $\Var(\hat{\rho})$ using the formula in equation (\ref{varrho2})
		\STATE \textbf{return} $\hat{\rho}$, $\sqrt{\Var\hat{\rho}}$
	\end{algorithmic}  
\end{algorithm}

Next, we generalize Algorithm 2 to the case where $X$ and $Y$ have $s$  and $t$ categories respectively, i.e, to estimate the polychoric correlation coefficient. As in the dichotomous case,  thresholds are  $a_i = \Phi^{-1}(P_{i\cdot}), i = 1, \dots, s$, $b_j = \Phi^{-1}(P_{\cdot j}), j = 1, \dots, r$.

Initially,  define
\begin{equation}
e_{x_{i}} \hat{=} E(Z_1| a_{i-1} < Z_1 \leq a_{i}) = \frac{\phi(a_{i-1}) - \phi(a_i)}{P_{i\cdot}} \label{iexim}
\end{equation}
for $i = 1, \dots, s$.

Conditional on $Z_1 = e_{x_{i}}$ and $b_{j-1} < Z_2  \leq b_{j}$,
\begin{equation}
e_{ij} = E(Z_2|b_{j-1} < Z_2  \leq b_{j}, Z_1 = e_{x_{i}}) \\ \label{eyim}
= \rho e_{x_i} +  \sqrt{1-\rho^2} \frac{\phi \left(e_{x_{i}}^{b_{j-1}} \right) - \phi \left(e_{x_{i}}^{b_{j}} \right)}{\Phi \left(e_{x_{i}}^{b_{j}} \right) - \Phi \left(e_{x_{i}}^{b_{j-1}} \right)}
\end{equation}
for $i = 1, \dots, s; j = 1, \dots, r$.

Then the conditional expected value of $Z_2$ given $Z_1 = e_{x_i}$ is
\begin{equation}
E_{Y_i} = E(Z_2 | Z_1 = e_{x_i}) = \sum_{j=1}^{r} \frac{P_{ij}}{P_{i\cdot}} e_{ij}  \label{eyxim}
\end{equation}
for $i = 1, \dots, s$.

Let frequency table $\hat{\mathbf{P}}_{s \times t}=(\hat{P}_{ij}), i=1,...,s; j=1,...,r$. The variance and covariances are listed in equation(\ref{var1}) and (\ref{cov}). Then the covariance matrix of vector $\mathbf{\hat{P}} $, by stacking $\hat{\mathbf{P}}$ by row, is given by 
\begin{equation}\label{cov2}
    \begin{split}
        \mathbf{B} &= (B_{ij})_{s\times r,s\times r}\\
        B_{ij} &= \begin{cases}
             \frac{1}{N}P_{i}(1-P_{i}),i=j,\\
          -\frac{1}{N}P_iP_j,i \neq j
         \end{cases}
    \end{split}
\end{equation}

For simplicity, we ignore the derivative brought by thresholds $b$ (This leads to a smaller final variance). The partial derivative matrix can be calculated as follows:
\begin{equation}\label{dera1}
    \begin{split} 
        \frac{\partial E_{Y_k}}{\partial P_{ij}} &=  
        \begin{cases}
            0, k \neq i\\
            \{(P_{k \cdot}-P_{kj})e_{kj} - \sum_{n=1,n\neq j}^r P_{kn}e_{kn}\}/P_{k \cdot}^2, k = i
        \end{cases}\\
    \end{split}
\end{equation}

Let $\mathbf{E_Y} = (E_{Y_i})_{s\times 1},i=1,...,s$, $\mathbf{D} = \frac{\partial{\mathbf{E_Y}}}{\partial{\mathbf{P}}} = (D_{ij})_{s,s\times r}$ is block diagonal matrix:
\begin{equation*}
         \left( 
            \begin{array}{cccc}
            \mathbf{D}_1  & 0  & \ldots & 0\\
            0 & \mathbf{D}_2 & \ldots & 0 \\
            \vdots & \vdots & \ddots & \vdots\\
            0 & 0 & \ldots & \mathbf{D}_s
            \end{array}
        \right)\\
\end{equation*}
where $\mathbf{D}_i$ is the non-zero elements derived by equation above which length $r$, and the covariance matrix is $\mathbf{\Sigma} = \mathbf{D}\mathbf{B}\mathbf{D'}$. Similar as  in estimating  tetrachoric correlation coefficient, $\hat{\rho}$ can be calculated with equation (\ref{rho2}). Variance of  $\hat{\rho}$ is calculated with (\ref{varrho2}). Standard error of $\hat{\rho}$ is obtained by taking the square root of its variance.

Then the predictors  $e_{x_i}$ are updated in the following two steps. First, to compute
\begin{equation}
\begin{split} 
E_{Y_{xj}} &= E(Z_2 |b_{j-1} < Z_2 \leq b_j) = \sum_{i=1}^{s} \frac{P_{ij}}{P_{\cdot j}} e_{ij}\\
e_{xij} &= E(Z_1|a_{i-1} < Z_1  \leq a_{i}, Z_2 =E_{Y_{xj}}) \\
&= \rho E_{Y_{xj}} + \sqrt{1-\rho^2} \frac{\phi \left(E_{Y_{xj}}^{a_{i-1}} \right) - \phi \left(E_{Y_{xj}}^{a_{i}} \right)}{\Phi \left(E_{Y_{xj}}^{a_{i}} \right) - \Phi \left(E_{Y_{xj}}^{a_{i-1}} \right)} \label{exyim}
\end{split}
\end{equation}
for $i = 1, \dots, s$ and $j = 1, \dots, r$. And then 
\begin{equation}
e_{x_i} = E(Z_1 | a_{i-1} < Z_1 \leq a_i) = \sum_{j=1}^{r} \frac{P_{ij}}{P_{i\cdot}} e_{xij}\label{exim}
\end{equation}
for $i = 1, \dots, s$.

The procedure is repeated until $\hat{\rho}$ converges. %

The details of the IRLS algorithm for estimating polychoric correlation coefficient are given in the following Algorithm \ref{poly},

\begin{algorithm}[H] 
	\renewcommand{\algorithmicrequire}{\textbf{Input:}}
	\renewcommand{\algorithmicensure}{\textbf{Output:}}
	\caption{IRLS method to compute polychoric correlation coefficient}
	\label{poly}
	\begin{algorithmic}[1]
		\REQUIRE observed ordinal data $y$ and $x$ or contingency table
		\ENSURE polychoric correlation of $Y$ and $X$, $\mbox{s.e.}(\hat{\rho})$
        \STATE Calculate the frequency table $\hat{\mathbf{P}}=(\hat{P}_{ij}),i=1,...,s;j=1,...,r$, the covariance $\mathbf{\hat{B}}$ using equation (\ref{cov2})
		\STATE Estimate the thresholds from the cumulative marginal proportions of $\hat{\mathbf{P}}$. $\hat{a}_i = \Phi^{-1}(\hat{P}_{i\cdot}), i = 1, \dots, s$, $\hat{b}_j = \Phi^{-1}(\hat{P}_{\cdot j}), j = 1, \dots, r$
		\STATE Initialization: \\
		$\hat{\rho}_0$ = Pearson correlation coefficient of $x$ and $y$, \\
		\STATE Initialize $\hat{e}_{x_{i}}$ using the formula in equation (\ref{iexim})for $i=1, \cdots, s$ 
		\STATE Set $\text{iter} = 0$, $\text{diff} = 1$, $n=100$ and $\epsilon = 1e{-8}$
		\WHILE{$\text{iter} < n ~\&~  \text{diff} > \epsilon$}
            \STATE Compute $e_{ij}$ for different categories using the formula in equation (\ref{eyim}) for $i=1, \cdots, s$ and $j=1, \cdots, r$
            \STATE Compute the derivation matrix $\mathbf{\hat{D}}$ using equation (\ref{dera1}), $\mathbf{\hat{\Sigma}} = \mathbf{\hat{D}}\mathbf{\hat{B}}\mathbf{\hat{D}'}$
			\STATE Compute $\hat{E}_{y_{i}}$ using the formula in equation (\ref{eyxim}) for $i=1, \cdots, s$
		    \STATE Compute $\hat{\rho}$ using formula in (\ref{rho2}) 
		    \STATE Compute $\text{diff} = \hat{\rho} - \hat{\rho}_0$, 
        
		    \STATE Compute 
		    $E_{Y_{xj}},e_{xij}$ using the  formula in equation (\ref{exyim}) 
		    for $i=1, \cdots, s$ and $j=1, \cdots, r$
		    \STATE Update $e_{x_i}$ using the formula in equation (\ref{exim}) for $i=1, \cdots, s$
		    \STATE Update $\hat{\rho}_0 = \hat{\rho}$ and iter = iter + 1
		
		\ENDWHILE
        \STATE Compute $\Var(\hat{\rho})$ using the formula in equation (\ref{varrho2})
		\STATE \textbf{return} estimate $\hat{\rho}$ and $\sqrt{\Var(\hat{\rho})}$ as the s.e. of $\hat{\rho}$.
	\end{algorithmic}  
\end{algorithm}

An R package IRLSpoly was developed to implement the Algorithm \ref{polyserial}, \ref{tetra}, \ref{poly}. It can be installed in R using the command {\textsf{install\textunderscore github(``encoreus/IRLSpoly")}}. 

\section{Simulation Study}

In this section, we conduct a series of simulation studies to compare the proposed algorithm with the standard maximum likelihood method. Following \cite{choi2011comparison}, the sample size is set to $N = 30, 50, 100, 500, 1000$. These numbers were chosen to reflect from small to moderate sample sizes that might be commonly encountered in social sciences. The population correlation coefficient is set to $\rho = 0, 0.2, 0.4, 0.6$, and $0.8$, ranging from null to moderate high. The number of categories for each ordinal variable are set to 2 (binary responses), 3, 5 and 7, i.e., $r = s = 2,3,5,7$ respectively. 

We generate data from a bivariate normal distribution with correlation $\rho$ and discretized them into categorical data, following the same procedure as described in Figure 1 in \cite{bollen1981pearson}, where the mean of the distribution is taken as a reference point and the variable is divided into equally-spaced intervals that move away from the mean of the distribution towards the extremes. Therefore, the distribution of the categorized data gets closer to normal as the number of categories is increased. 

To measure the performance of the two methods, we use the following criteria:
\begin{itemize}
\item $\mbox{MEAN} = \sum_{i=1}^{n}{\hat{\rho}}/{n},$\\
which is the mean value of the estimates, where 
$n$ is the replication number attempted (i.e., $n = 1,000$)
\item $\mbox{MRB} = \sum_{i=1}^{\text{n}}\{(\hat{\rho_i} - \rho)/\rho\} / \text{n}$, \\
which is the mean relative bias (MRB) to evaluate the bias of the estimators, where $\hat{\rho}$ is the estimator in the $i$th replication, $\rho$ is the true value. The general form is given in \cite{bandalos2006use}.\\
$\mbox{MB} = \sum_{i=1}^{\text{n}}(\hat{\rho_i} - \rho) / \text{n}$. \\
In case of $\rho = 0$, mean bias (MB) is used instead of MRB, to avoid the issue of dividing by zero.
\item $\mbox{RMSE} = \sqrt{\sum_{i=1}^{\text{n}}(\hat{\rho} - \rho)^2 / \text{n}}$, \\
which is the root mean squared error (RMSE), to evaluate the variability of the estimators.
\item $\mbox{SD} = \sqrt{\sum_{i=1}^{\text{n}}(\hat{\rho} - MEAN)^2 / \text{n}}$,\\
which is the standard deviation of mean values (SD), to evaluate the variability of the estimators.
\item $\mbox{MSD} = \sum_{i=1}^n\sqrt{Var(\hat{\rho})}/n$,\\
which is the mean value of the estimates of the variance of $\rho$. We compare it with SD to verify the accuracy of the algorithm variance calculation formula.

\end{itemize}


For polyserial correlation, the results with true value of $\rho = 0, 0.2, 0.4, 0.6, 0.8$ are shown in Table \ref{tab1}. Function  \texttt{polyserial} defined in the \texttt{polycor} package in R by \cite{fox2010polycor} is invoked to obtain the MLEs.  
\setlength{\tabcolsep}{0.8mm}{\label{simu5}
\begin{table}[H]\large
  \centering
  \fontsize{5}{5}\selectfont  
  \begin{threeparttable}  
    \begin{tabular}{ccccccccccccccccccc}
    \toprule  
    \multicolumn{1}{l}{} &\multicolumn{5}{c}{IRLS} &\multicolumn{4}{c}{ML method}&\multicolumn{5}{c}{IRLS}&\multicolumn{4}{c}{ML method}\cr
    \cmidrule(lr){2-6}\cmidrule(lr){7-10}\cmidrule(lr){11-15}\cmidrule(lr){16-19}
    N            &MEAN         &SD        &MRB      &RMSE      &MSD       &MEAN         &SD        &MRB      &RMSE  &MEAN         &SD        &MRB      &RMSE      &MSD    &MEAN         &SD        &MRB      &RMSE      \cr 
    \cmidrule(lr){1-1}\cmidrule(lr){2-10} \cmidrule(lr){11-19}
    & \multicolumn{9}{c}{$\rho = 0, s = 2$} &\multicolumn{9}{c}{$\rho = 0, s  = 3$} \cr
    \midrule 
30 &0.0012 &0.2307 &0.0012 &0.2307 &0.2271 &-0.0002 &0.2399 &-0.0002 &0.2398 &0.0019 &0.2062 &0.0019 &0.2061 &0.2036 &0.0007 &0.2144 &0.0007 &0.2142\cr 
50 &-0.0023 &0.1773 &-0.0023 &0.1773 &0.1765 &-0.0023 &0.1811 &-0.0023 &0.1810 &0.0031 &0.1579 &0.0031 &0.1579 &0.1582 &0.0032 &0.1614 &0.0032 &0.1614\cr 
100 &-0.0004 &0.1256 &-0.0004 &0.1255 &0.1251 &0.0000 &0.1266 &0.0001 &0.1265 &0.0032 &0.1100 &0.0033 &0.1100 &0.1120 &0.0037 &0.1111 &0.0037 &0.1111 \cr 
500 &-0.0022 &0.0558 &-0.0022 &0.0558 &0.0560 &-0.0022 &0.0560 &-0.0022 &0.0560 &-0.0005 &0.0505 &-0.0005 &0.0505 &0.0502 &-0.0005 &0.0506 &-0.0005 &0.0506\cr 
1000 &-0.0016 &0.0389 &-0.0016 &0.0390 &0.0396 &-0.0016 &0.0390 &-0.0016 &0.0390 &-0.0008 &0.0353 &-0.0008 &0.0353 &0.0355 &-0.0007 &0.0354 &-0.0007 &0.0364\cr 
    \cmidrule(lr){1-10} \cmidrule(lr){11-19}
    & \multicolumn{9}{c}{$\rho = 0, s = 5$} &\multicolumn{9}{c}{$\rho = 0, s = 7$} \cr
    \midrule
30 &0.0015 &0.1951 &0.0015 &0.1950 &0.1912 &0.0004 &0.2033 &0.0004 &0.2032 &0.0003 &0.1907 &0.0003 &0.1906 &0.1872 &-0.0008 &0.1995 &-0.0008 &0.1994 \cr 
50 &0.0017 &0.1486 &0.0017 &0.1485 &0.1486 &0.0013 &0.1518 &0.0013 &0.1518 &-0.0002 &0.1464 &-0.0002 &0.1464 &0.1455 &-0.0004 &0.1501 &-0.0004 &0.1501\cr 
100 &0.0011 &0.1057 &0.0011 &0.1056 &0.1053 &0.0014 &0.1066 &0.0014 &0.1066 &-0.0007 &0.1033 &-0.0007 &0.1033 &0.1032 &-0.0004 &0.1043 &-0.0004 &0.1042\cr 
500 &-0.0007 &0.0475 &-0.0007 &0.0475 &0.0472 &-0.0007 &0.0477 &-0.0007 &0.0477 &-0.0013 &0.0464 &-0.0013 &0.0464 &0.0462 &-0.0012 &0.0466 &-0.0012 &0.0466\cr 
1000 &-0.0003 &0.0336 &-0.0003 &0.0336 &0.0334 &-0.0003 &0.0336 &-0.0003 &0.0336 &-0.0008 &0.0330 &-0.0008 &0.033 &0.0327 &-0.0008 &0.033 &-0.0008 &0.0330 \cr 
    \cmidrule(lr){1-10} \cmidrule(lr){11-19}
    & \multicolumn{9}{c}{$\rho = 0.2, s = 2$} &\multicolumn{9}{c}{$\rho = 0.2, s = 3$} \cr
    \midrule
30 &0.1987 &0.2293 &-0.0013 &0.2292 &0.2239 &0.2032 &0.2325 &0.0032 &0.2324 &0.1973 &0.2032 &-0.0027 &0.2032 &0.2006 &0.2019 &0.2060 &0.0019 &0.2059\cr 
50 &0.2001 &0.1763 &0.0001 &0.1762 &0.1741 &0.2028 &0.1760 &0.0028 &0.1760 &0.2010 &0.1551 &0.0010 &0.1550 &0.1558 &0.2038 &0.1547 &0.0038 &0.1547\cr 
100 &0.2014 &0.1245 &0.0014 &0.1244 &0.1234 &0.2027 &0.1235 &0.0027 &0.1235 &0.2014 &0.1080 &0.0014 &0.1080 &0.1104 &0.2030 &0.1075 &0.0030 &0.1075\cr 
500 &0.1981 &0.0562 &-0.0019 &0.0562 &0.0553 &0.1985 &0.0554 &-0.0015 &0.0554 &0.1988 &0.0497 &-0.0012 &0.0497 &0.0495 &0.1991 &0.0488 &-0.0009 &0.0488\cr 
1000 &0.1989 &0.0380 &-0.0011 &0.0380 &0.0391 &0.1992 &0.0378 &-0.0008 &0.0377 &0.1988 &0.0346 &-0.0012 &0.0346 &0.0350 &0.1992 &0.0343 &-0.0008 &0.0343\cr
    \cmidrule(lr){1-10} \cmidrule(lr){11-19}
    & \multicolumn{9}{c}{$\rho = 0.2, s = 5$} &\multicolumn{9}{c}{$\rho = 0.2, s = 7$} \cr
    \midrule
30 &0.1957 &0.1965 &-0.0043 &0.1964 &0.1880 &0.2001 &0.1981 &0.0001 &0.1980 &0.1947 &0.1906 &-0.0053 &0.1906 &0.1839 &0.2001 &0.1929 &0.0001 &0.1928\cr 
50 &0.1987 &0.1472 &-0.0013 &0.1472 &0.1462 &0.2013 &0.1461 &0.0013 &0.1460 &0.1973 &0.1427 &-0.0027 &0.1426 &0.1431 &0.2006 &0.1420 &0.0006 &0.1419\cr 
100 &0.1997 &0.1029 &-0.0003 &0.1028 &0.1036 &0.2014 &0.1017 &0.0014 &0.1017 &0.1991 &0.0998 &-0.0009 &0.0997 &0.1014 &0.2010 &0.0986 &0.0010 &0.0986\cr 
500 &0.1985 &0.0465 &-0.0015 &0.0465 &0.0464 &0.1989 &0.0457 &-0.0011 &0.0457 &0.1992 &0.0461 &-0.0008 &0.0461 &0.0454 &0.1997 &0.0452 &-0.0003 &0.0452\cr 
1000 &0.1987 &0.0323 &-0.0013 &0.0323 &0.0328 &0.1991 &0.0319 &-0.0009 &0.0319 &0.1993 &0.0319 &-0.0007 &0.0319 &0.0321 &0.1997 &0.0315 &-0.0003 &0.0314\cr 
    \cmidrule(lr){1-10} \cmidrule(lr){11-19}
    & \multicolumn{9}{c}{$\rho = 0.4, s = 2$} &\multicolumn{9}{c}{$\rho = 0.4, s  = 3$} \cr
    \midrule
30 &0.3865 &0.2170 &-0.0135 &0.2173 &0.2154 &0.3976 &0.2083 &-0.0024 &0.2082 &0.3952 &0.1881 &-0.0048 &0.1880 &0.1915 &0.4043 &0.1780 &0.0043 &0.1779\cr 
50 &0.3949 &0.1636 &-0.0051 &0.1636 &0.1673 &0.4009 &0.1545 &0.0009 &0.1544 &0.3951 &0.1448 &-0.0049 &0.1448 &0.1488 &0.4014 &0.1345 &0.0014 &0.1344\cr 
100 &0.3976 &0.1104 &-0.0024 &0.1103 &0.1186 &0.4003 &0.1039 &0.0003 &0.1038 &0.3988 &0.1064 &-0.0012 &0.1064 &0.1053 &0.4027 &0.0974 &0.0027 &0.0974\cr 
500 &0.3986 &0.0524 &-0.0014 &0.0524 &0.0531 &0.3994 &0.0494 &-0.0006 &0.0494 &0.3988 &0.0479 &-0.0012 &0.0479 &0.0472 &0.3998 &0.0435 &-0.0002 &0.0435\cr 
1000 &0.3986 &0.0370 &-0.0014 &0.0370 &0.0376 &0.3992 &0.0354 &-0.0008 &0.0354 &0.4000 &0.0325 &0.0000 &0.0325 &0.0334 &0.4003 &0.0292 &0.0003 &0.0292 \cr 
    \cmidrule(lr){1-10} \cmidrule(lr){11-19}
    & \multicolumn{9}{c}{$\rho = 0.4, s = 5$} &\multicolumn{9}{c}{$\rho = 0.4, s = 7$} \cr
    \midrule
30 &0.3917 &0.1883 &-0.0083 &0.1884 &0.1799 &0.4026 &0.1736 &0.0026 &0.1735 &0.3990 &0.1743 &-0.0010 &0.1742 &0.1742 &0.4104 &0.1626 &0.0104 &0.1628\cr 
50 &0.3970 &0.1394 &-0.0030 &0.1394 &0.1391 &0.4031 &0.1266 &0.0031 &0.1265 &0.3968 &0.1359 &-0.0032 &0.1359 &0.1367 &0.4043 &0.1244 &0.0043 &0.1244\cr 
100 &0.4030 &0.1019 &0.0030 &0.1019 &0.0981 &0.4059 &0.0912 &0.0059 &0.0913 &0.3971 &0.1010 &-0.0029 &0.1010 &0.0959 &0.3995 &0.0909 &-0.0005 &0.0909\cr 
500 &0.4008 &0.0468 &0.0008 &0.0468 &0.0440 &0.4011 &0.0416 &0.0011 &0.0416 &0.3979 &0.0432 &-0.0021 &0.0432 &0.0430 &0.3995 &0.0378 &-0.0005 &0.0378\cr 
1000 &0.3987 &0.0302 &-0.0013 &0.0302 &0.4713 &0.3998 &0.0277 &-0.0002 &0.0277 &0.3993 &0.0310 &-0.0007 &0.0310 &0.3902 &0.4001 &0.0275 &0.0001 &0.0275 \cr
    \cmidrule(lr){1-10} \cmidrule(lr){11-19}
    & \multicolumn{9}{c}{$\rho = 0.6, s = 2$} &\multicolumn{9}{c}{$\rho = 0.6, s = 3$} \cr
    \midrule
30 &0.5860 &0.1982 &-0.0140 &0.1984 &0.1993 &0.6044 &0.1681 &0.0044 &0.1680 &0.5966 &0.1875 &-0.0034 &0.1873 &0.1743 &0.6081 &0.1571 &0.0081 &0.1571\cr 
50 &0.5936 &0.1492 &-0.0064 &0.1492 &0.1549 &0.6028 &0.1245 &0.0028 &0.1245 &0.5958 &0.1371 &-0.0042 &0.1371 &0.1358 &0.6048 &0.1089 &0.0048 &0.1090 \cr 
100 &0.5971 &0.1043 &-0.0029 &0.1043 &0.1098 &0.6008 &0.0851 &0.0008 &0.0851 &0.5951 &0.0966 &-0.0049 &0.0967 &0.0965 &0.6014 &0.0756 &0.0014 &0.0756\cr 
500 &0.5991 &0.0490 &-0.0009 &0.0490 &0.0492 &0.6002 &0.0404 &0.0002 &0.0404 &0.5980 &0.0426 &-0.0020 &0.0426 &0.0432 &0.5995 &0.0331 &-0.0005 &0.0330\cr 
1000 &0.5993 &0.0345 &-0.0007 &0.0345 &0.0348 &0.5998 &0.0274 &-0.0002 &0.0274 &0.6002 &0.0311 &0.0002 &0.0311 &0.0305 &0.6001 &0.0233 &0.0001 &0.0233\cr
    \cmidrule(lr){1-10} \cmidrule(lr){11-19}
    & \multicolumn{9}{c}{$\rho = 0.6, s = 5$} &\multicolumn{9}{c}{$\rho = 0.6, s = 7$} \cr
    \midrule
30 &0.5842 &0.1731 &-0.0158 &0.1738 &0.1688 &0.6028 &0.1359 &0.0028 &0.1359 &0.5927 &0.1591 &-0.0073 &0.1592 &0.1693 &0.6096 &0.1218 &0.0096 &0.1221\cr 
50 &0.5944 &0.1290 &-0.0056 &0.1291 &0.1289 &0.6042 &0.0978 &0.0042 &0.0978 &0.5946 &0.1288 &-0.0054 &0.1288 &0.1540 &0.6057 &0.0965 &0.0057 &0.0967\cr 
100 &0.5996 &0.0947 &-0.0004 &0.0947 &0.0885 &0.6039 &0.0702 &0.0039 &0.0703 &0.5963 &0.0952 &-0.0037 &0.0952 &0.0858 &0.5998 &0.0704 &-0.0002 &0.0704\cr 
500 &0.6001 &0.0431 &0.0001 &0.0431 &0.0397 &0.6007 &0.0317 &0.0007 &0.0317 &0.5969 &0.0410 &-0.0031 &0.0411 &0.0386 &0.5991 &0.0291 &-0.0009 &0.0291 \cr 
1000 &0.5987 &0.0290 &-0.0013 &0.0290 &0.0281 &0.5995 &0.0209 &-0.0005 &0.0209 &0.5998 &0.0289 &-0.0002 &0.0289 &0.0272 &0.6001 &0.0208 &0.0001 &0.0208  \cr 
    \cmidrule(lr){1-10} \cmidrule(lr){11-19}
    & \multicolumn{9}{c}{$\rho = 0.8, s = 2$} &\multicolumn{9}{c}{$\rho = 0.8, s = 3$} \cr
    \midrule
30 &0.7783 &0.1599 &-0.0217 &0.1613 &0.1764 &0.8111 &0.1151 &0.8111 &0.1156 &0.7776 &0.1486 &-0.0224 &0.1502 &0.1513 &0.8107 &0.0970 &0.8107 &0.0976\cr 
50 &0.7899 &0.1290 &-0.0101 &0.1293 &0.1359 &0.8060 &0.0922 &0.806 &0.0924 &0.7902 &0.1189 &-0.0098 &0.1192 &0.1162 &0.8038 &0.0725 &0.8038 &0.0726\cr 
100 &0.7974 &0.0914 &-0.0026 &0.0914 &0.0961 &0.8029 &0.0548 &0.0029 &0.0549 &0.7911 &0.0823 &-0.0089 &0.0827 &0.0827 &0.8012 &0.0448 &0.0012 &0.0448 \cr 
500 &0.7999 &0.0414 &-0.0001 &0.0414 &0.0431 &0.8015 &0.0250 &0.0015 &0.0250 &0.7985 &0.0377 &-0.0015 &0.0377 &0.0419 &0.8004 &0.0197 &0.0004 &0.0197 \cr 
1000 &0.7983 &0.0299 &-0.0017 &0.0299 &0.0305 &0.7988 &0.0166 &-0.0012 &0.0166 &0.7998 &0.0280 &-0.0002 &0.0280 &0.0261 &0.8000 &0.0142 &0.0000 &0.0142 \cr
    \cmidrule(lr){1-10} \cmidrule(lr){11-19}
    & \multicolumn{9}{c}{$\rho = 0.8, s = 5$} &\multicolumn{9}{c}{$\rho = 0.8, s = 7$} \cr
    \midrule
30 &0.7745 &0.1441 &-0.0255 &0.1463 &0.1317 &0.8094 &0.0787 &0.0094 &0.0792 &0.7689 &0.1504 &-0.0311 &0.1535 &0.1223 &0.8092 &0.0721 &0.0092 &0.0726\cr 
50 &0.7875 &0.1164 &-0.0125 &0.1170 &0.1020 &0.8054 &0.0578 &0.0054 &0.0581 &0.7878 &0.1141 &-0.0122 &0.1147 &0.0956 &0.8056 &0.0559 &0.0056 &0.0562\cr 
100 &0.7984 &0.0869 &-0.0016 &0.0869 &0.0721 &0.8037 &0.0410 &0.0037 &0.0411 &0.7969 &0.0850 &-0.0031 &0.0850 &0.0685 &0.8026 &0.0400 &0.0026 &0.0401\cr 
500 &0.8000 &0.0386 &0.0000 &0.0386 &0.0326 &0.8008 &0.0182 &0.0008 &0.0182 &0.7964 &0.0379 &-0.0036 &0.0380 &0.0311 &0.7992 &0.0168 &-0.0008 &0.0168\cr 
1000 &0.7989 &0.0271 &-0.0011 &0.0271 &0.0231 &0.7999 &0.0124 &-0.0001 &0.0124 &0.7999 &0.0264 &-0.0001 &0.0264 &0.0219 &0.8002 &0.0120 &0.0002 &0.0120\cr 
    \bottomrule  
    \end{tabular}  
    \caption{Simulation results of polyserial correlation comparing  the IRLS algorithm with the ML method}\label{tab1}
    \end{threeparttable}  
\end{table} }
It can be seen that the results of the two methods are similar, especially with $\rho = 0, 0.2, 0.4$. When $\rho = 0.6, 0.8$, for small sample size such as $N = 30, 50$ the bias of the new method is slightly larger than that of the ML method. However, as the sample size increase, it performs as well as the ML method. In addition, the mean value of the standard deviation  calculated by the new method (MSD) is very close to that of Monte Carlo (SD). 

The slightly larger biases from the IRLS are mainly due to the fact that only  the means of the data, instead of the whole data, are utilized in estimating polyserial correlations. On the other hand, the advantages of the proposed method for polyserial correlation include: 1) it is obtained from data summaries. If only summaries were reported, for example, in a report or a paper,  the proposed method can be applied to estimate the polyserial correlation, but the traditional ML method cannot; 2) In the simulation, the ML may not be able to calculate because of the small $n$ and $s$ (leading a null category), while the proposed method still works.

For polychoric correlation simulations, a quick two step maximum likelihood(\cite{olsson1979maximum}) procedure is adopted in calling \texttt{polychor} function in the \texttt{polycor} package of R. The regular ML method in which the thresholds are estimated simultaneously is too slow to run the simulations. We use the same settings as in simulations for polyserial correlations.

\setlength{\tabcolsep}{0.8mm}{\label{simu5}
\begin{table}[H]\large
  \centering
  \fontsize{5}{5}\selectfont  
  \begin{threeparttable}  
    \begin{tabular}{ccccccccccccccccccc}
    \toprule  
    \multicolumn{1}{l}{} &\multicolumn{5}{c}{IRLS} &\multicolumn{4}{c}{ML method}&\multicolumn{5}{c}{IRLS}&\multicolumn{4}{c}{ML method}\cr
    \cmidrule(lr){2-6}\cmidrule(lr){7-10}\cmidrule(lr){11-15}\cmidrule(lr){16-19}
    N            &MEAN         &SD        &MRB        &RMSE      &MSD       &MEAN         &SD        &MRB      &RMSE  &MEAN         &SD        &MRB      &RMSE        &MSD    &MEAN         &SD        &MRB      &RMSE      \cr 
    \cmidrule(lr){1-1}\cmidrule(lr){2-10} \cmidrule(lr){11-19}
    & \multicolumn{9}{c}{$\rho = 0, s = r = 2$} &\multicolumn{9}{c}{$\rho = 0, s = r = 3$} \cr
    \midrule 
30 &-0.0034 &0.2863 &-0.0034 &0.2862 &0.1755 &0.0001 &0.2866 &0.0001 &0.2864 &-0.0063 &0.2390 &-0.0063 &0.2390 &0.1707 &-0.0033 &0.2373 &-0.0033 &0.2374\cr 
50 &-0.0007 &0.2249 &-0.0007 &0.2248 &0.1382 &0.0012 &0.2252 &0.0012
&0.2248 &0.0075 &0.1811 &0.0075 &0.1811 &0.1368 &-0.0053 &0.1801 &-0.0053 &0.1801\cr 
100 &0.0032 &0.1548 &0.0032 &0.1547 &0.0990 &-0.0001 &0.1564 &-0.0001 &0.1563 &-0.0068 &0.1280 &-0.0068 &0.1282 &0.0984 &0.0010 &0.1273 &0.0010 &0.1272 \cr 
500 &0.0006 &0.0701 &0.0006 &0.0700 &0.0446 &-0.0005 &0.0706 &-0.0005 &0.0705 &0.0021 &0.0556 &0.0021 &0.0556 &0.0446 &-0.0009 &0.0588 &-0.0009 &0.0588\cr 
1000 &0.0011 &0.0506 &0.0011 &0.0506 &0.0316 &0.0007 &0.0493 &0.0007 &0.0493 &-0.0006 &0.0384 &-0.0006 &0.0384 &0.0316 &0.0006 &0.0396 &0.0006 &0.0396\cr 
    \cmidrule(lr){1-10} \cmidrule(lr){11-19}
    & \multicolumn{9}{c}{$\rho = 0, s = r = 5$} &\multicolumn{9}{c}{$\rho = 0, s = r = 7$} \cr
    \midrule
30 &-0.0055 &0.2372 &-0.0055 &0.2371 &0.1498 &0.0045 &0.2026 &0.0045 &0.2026
&0.0009 &0.2439 &0.0009 &0.2436 &0.1298 &-0.0086 &0.1998 &-0.0086 &0.1999\cr 
50 &0.0006 &0.1692 &0.0006 &0.1691 &0.1290 &0.0078 &0.1593 &0.0078 &0.1594 &0.0055 &0.1799 &0.0055 &0.1798 &0.1176 &-0.0083 &0.151 &-0.0083 &0.1512\cr 
100 &0.0034 &0.1159 &0.0034 &0.1159 &0.0960 &0.0004 &0.1111 &0.0004 &0.1111 &-0.0031 &0.1108 &-0.0031 &0.1108 &0.0930 &-0.0020 &0.1068 &-0.0020 &0.1068\cr 
500 &0.0007 &0.0483 &0.0007 &0.0483 &0.0444 &-0.0009 &0.0487 &-0.0009 &0.0487 &-0.0014 &0.0472 &-0.0014 &0.0472 &0.0442 &-0.0012 &0.0486 &-0.0012 &0.0486 \cr 
1000 &-0.0003 &0.0351 &-0.0003 &0.0351 &0.0315 &0.0001 &0.0339 &0.0001 &0.0339 &0.0013 &0.0340 &0.0013 &0.0340 &0.0314 &-0.0008 &0.0341 &-0.0008 &0.0341 \cr 
    \cmidrule(lr){1-10} \cmidrule(lr){11-19}
    & \multicolumn{9}{c}{$\rho = 0.2, s = r = 2$} &\multicolumn{9}{c}{$\rho = 0.2, s = r = 3$} \cr
    \midrule
30 &0.1949 &0.2726 &-0.0051 &0.2725 &0.1727 &0.2002 &0. 2760 &0.001 &0.2758 &0.1973 &0.2334 &-0.0027 &0.2333 &0.1668 &0.2059 &0.2269 &0.0297 &0.2269 \cr 
50 &0.1974 &0.2147 &-0.0026 &0.2147 &0.1358 &0.2050 &0.2183 &0.0249 &0.2182 &0.2047 &0.1760 &0.0047 &0.1760 &0.1336 &0.1996 &0.1757 &-0.0018 &0.1756  \cr 
100 &0.2014 &0.1493 &0.0014 &0.1493 &0.0972 &0.1991 &0.1500 &-0.0046 &0.15 &0.1919 &0.1238 &-0.0081 &0.1240 &0.0965 &0.1998 &0.1220 &-0.0011 &0.1220\cr 
500 &0.2011 &0.0663 &0.0011 &0.0663 &0.0438 &0.1983 &0.0672 &-0.0085 &0.0672 &0.2015 &0.0528 &0.0015 &0.0528 &0.0436 &0.2000 &0.0549 &0.0000 &0.0549 \cr 
1000 &0.1997 &0.0481 &-0.0003 &0.0480 &0.0310 &0.2007 &0.0494 &0.0033 &0.0494 &0.1987 &0.0372 &-0.0013 &0.0372 &0.0309 &0.1997 &0.0389 &-0.0015 &0.0389\cr
    \cmidrule(lr){1-10} \cmidrule(lr){11-19}
    & \multicolumn{9}{c}{$\rho = 0.2, s = r = 5$} &\multicolumn{9}{c}{$\rho = 0.2, s = r = 7$} \cr
    \midrule
30 &0.2178 &0.2311 &0.0178 &0.2316 &0.1467 &0.1958 &0.1976 &-0.0211 &0.2063 &0.1998 &0.2290 &-0.0002 &0.2288 &0.1274 &0.2026 &0.1917 &0.0131 &0.1916\cr 
50 &0.2112 &0.1688 &0.0112 &0.1691 &0.1249 &0.2059 &0.1541 &0.0294 &0.1542 &0.2148 &0.1709 &0.0148 &0.1714 &0.1161 &0.1952 &0.1526 &-0.0239 &0.1526\cr 
100 &0.2066 &0.1115 &0.0066 &0.1116 &0.0940 &0.1972 &0.1064 &-0.0138 &0.1063 &0.2000 &0.1142 &0.0000 &0.1141 &0.0911 &0.2009 &0.1049 &0.0044 &0.1049\cr 
500 &0.1999 &0.0488 &-0.0001 &0.0488 &0.0435 &0.2006 &0.0478 &0.0032 &0.0478 &0.2008 &0.0440 &0.0008 &0.0440 &0.0433 &0.2005 &0.0444 &0.0027 &0.0444 \cr 
1000 &0.2002 &0.0335 &0.0002 &0.0335 &0.0309 &0.1998 &0.0338 &-0.0011 &0.0338 &0.1990 &0.0318 &-0.0010 &0.0318 &0.0308 &0.2003 &0.0328 &0.0014 &0.0328\cr 
    \cmidrule(lr){1-10} \cmidrule(lr){11-19}
    & \multicolumn{9}{c}{$\rho = 0.4, s = r = 2$} &\multicolumn{9}{c}{$\rho = 0.4, s = r = 3$} \cr
    \midrule
30 &0.3896 &0.2471 &-0.0104 &0.2472 &0.1633 &0.4144 &0.2431 &0.0361 &0.2434 &0.4035 &0.2112 &0.0035 &0.2112 &0.1551 &0.3975 &0.2049 &-0.0063 &0.2048\cr 
50 &0.3936 &0.1944 &-0.0064 &0.1944 &0.1283 &0.4050 &0.1930 &0.0125 &0.1929 &0.4060 &0.1546 &0.0060 &0.1546 &0.1244 &0.4024 &0.1581 &0.0059 &0.1580\cr 
100 &0.4006 &0.1367 &0.0006 &0.1367 &0.0916 &0.4059 &0.1386 &0.0147 &0.1387 &0.3936 &0.1108 &-0.0064 &0.1109 &0.0901 &0.4031 &0.1125 &0.0077 &0.1125\cr 
500 &0.3973 &0.0607 &-0.0027 &0.0608 &0.0414 &0.4018 &0.0651 &0.0045 &0.0651 &0.3983 &0.0471 &-0.0017 &0.0471 &0.0408 &0.4022 &0.0488 &0.0055 &0.0488\cr 
1000&0.3969 &0.0433 &-0.0031 &0.0433 &0.0293 &0.4009 &0.0454 &0.0022 &0.0454 &0.3968 &0.0335 &-0.0032 &0.0336 &0.0289 &0.4014 &0.0342 &0.0034 &0.0342\cr 
    \cmidrule(lr){1-10} \cmidrule(lr){11-19}
    & \multicolumn{9}{c}{$\rho = 0.4, s = r = 5$} &\multicolumn{9}{c}{$\rho = 0.4, s = r = 7$} \cr
    \midrule
30 &0.4214 &0.2045 &0.0214 &0.2055 &0.1375 &0.4055 &0.1777 &0.0136 &0.1777 &0.3970 &0.2065 &-0.0030 &0.2063 &0.1218 &0.3967 &0.1742 &-0.0081 &0.1742\cr 
50 &0.4166 &0.1483 &0.0166 &0.1491 &0.1163 &0.4005 &0.1354 &0.0012 &0.1353 &0.4203 &0.1552 &0.0203 &0.1564 &0.1085 &0.3976 &0.1330 &-0.0059 &0.1329 \cr 
100 &0.4093 &0.0983 &0.0093 &0.0987 &0.0874 &0.4009 &0.0965 &0.0022 &0.0964 &0.4040 &0.1019 &0.0040 &0.1019 &0.0852 & 0.4008 &0.0881 &0.0020 &0.0881\cr 
500 &0.3994 &0.0439 &-0.0006 &0.0439 &0.0405 &0.4009 &0.0413 &0.0022 &0.0413 &0.3997 &0.0394 &-0.0003 &0.0394 &0.0404 &0.3996 &0.0411 &-0.0009 &0.0411\cr 
1000 &0.3988 &0.0296 &-0.0012 &0.0296 &0.0288 &0.4009 &0.0304 &0.0024 &0.0304 &0.3988 &0.0280 &-0.0012 &0.0280 &0.0287 &0.4006 &0.0284 &0.0016 &0.0283\cr
    \cmidrule(lr){1-10} \cmidrule(lr){11-19}
    & \multicolumn{9}{c}{$\rho = 0.6, s = r = 2$} &\multicolumn{9}{c}{$\rho = 0.6, s = r = 3$} \cr
    \midrule
30 &0.5753 &0.2039 &-0.0247 &0.2053 &0.1470 &0.5911 &0.2009 &-0.0148 &0.2010 &0.6097 &0.1656 &0.0097 &0.1658 &0.1334 &0.607 &0.1613 &0.0116 &0.1614\cr 
50 &0.5837 &0.1599 &-0.0163 &0.1606 &0.1149 &0.5973 &0.1614 &-0.0045 &0.1614 &0.6082 &0.1289 &0.0082 &0.1291 &0.1068 &0.6075 &0.1300 &0.0124 &0.1301\cr 
100 &0.5921 &0.1097 &-0.0079 &0.1099 &0.0819 &0.5990 &0.1211 &-0.0016 &0.1210 &0.5929 &0.0892 &-0.0071 &0.0895 &0.0784 &0.6042 &0.0875 &0.0070 &0.0875\cr 
500 &0.5901 &0.0484 &-0.0099 &0.0494 &0.0371 &0.6008 &0.0500 &0.0013 &0.05 &0.5927 &0.0370 &-0.0073 &0.0377 &0.0356 &0.601 &0.0385 &0.0017 &0.0384\cr 
1000 &0.5905 &0.0350 &-0.0095 &0.0362 &0.0262 &0.6005 &0.0359 &0.0009 &0.0359 &0.5925 &0.0262 &-0.0075 &0.0273 &0.0252 &0.6007 &0.0283 &0.0011 &0.0283\cr
    \cmidrule(lr){1-10} \cmidrule(lr){11-19}
    & \multicolumn{9}{c}{$\rho = 0.6, s = r = 5$} &\multicolumn{9}{c}{$\rho = 0.6, s = r = 7$} \cr
    \midrule
30 &0.6139 &0.1623 &0.0139 &0.1629 &0.1206 &0.5984 &0.1406 &-0.0026 &0.1405 &0.6019 &0.1787 &0.0019 &0.1786 &0.1101 &0.5902 &0.1380 &-0.0163 &0.1383\cr 
50 &0.6197 &0.1169 &0.0197 &0.1185 &0.1001 &0.5984 &0.1094 &-0.0026 &0.1094 &0.6252 &0.1191 &0.0252 &0.1217 &0.0941 &0.6005 &0.1028 &0.0008 &0.1027 \cr 
100 &0.6077 &0.0764 &0.0077 &0.0767 &0.0752 &0.6012 &0.0755 &0.0021 &0.0755 &0.6079 &0.0795 &0.0079 &0.0798 &0.0737 &0.5997 &0.0719 &-0.0005 &0.0718\cr 
500 &0.5969 &0.0336 &-0.0031 &0.0338 &0.0351 &0.5988 &0.0342 &-0.0020 &0.0342 &0.5983 &0.0308 &-0.0017 &0.0308 &0.0350 &0.5999 &0.0320 &-0.0002 &0.0320\cr 
1000 &0.5956 &0.0220 &-0.0044 &0.0224 &0.0249 &0.6005 &0.0238 &0.0008 &0.0238 &0.5976 &0.0219 &-0.0024 &0.0220 &0.0248 &0.5998 &0.0221 &-0.0003 &0.0221\cr 
    \cmidrule(lr){1-10} \cmidrule(lr){11-19}
    & \multicolumn{9}{c}{$\rho = 0.8, s = r = 2$} &\multicolumn{9}{c}{$\rho = 0.8, s = r = 3$} \cr
    \midrule
30 &0.7616 &0.1434 &-0.0384 &0.1484 &0.1211 &0.7619 &0.1399 &-0.0477 &0.1449 &0.7901 &0.1016 &-0.0099 &0.1021 &0.1027 &0.7955 &0.1066 &-0.0056 &0.1066\cr 
50 &0.7773 &0.1049 &-0.0227 &0.1073 &0.0921 &0.7815 &0.1050 &-0.0231 &0.1065 &0.8032 &0.0814 &0.0032 &0.0814 &0.0786 &0.7984 &0.0813 &-0.0020 &0.0812\cr 
100 &0.7833 &0.0724 &-0.0167 &0.0743 &0.0654 &0.7951 &0.0730 &-0.0062 &0.0731 &0.7957 &0.0569 &-0.0043 &0.0570 &0.0579 &0.8024 &0.0549 &0.0030 &0.0549 \cr 
500 &0.7809 &0.0313 &-0.0191 &0.0367 &0.0298 &0.7988 &0.0330 &-0.0016 &0.033 &0.7895 &0.0248 &-0.0105 &0.0269 &0.0267 &0.8003 &0.0257 &0.0003 &0.0257\cr 
1000 &0.7810 &0.0224 &-0.0190 &0.0294 &0.0211 &0.7989 &0.0232 &-0.0014 &0.0232 &0.7885 &0.0173 &-0.0115 &0.0208 &0.0190 &0.8006 &0.0178 &0.0008 &0.0178\cr
    \cmidrule(lr){1-10} \cmidrule(lr){11-19}
    & \multicolumn{9}{c}{$\rho = 0.8, s = r = 5$} &\multicolumn{9}{c}{$\rho = 0.8, s = r = 7$} \cr
    \midrule
30 &0.8147 &0.0919 &0.0147 &0.0930 &0.0898 &0.8026 &0.0850 &0.0032 &0.0850 &0.7966 &0.1102 &-0.0034 &0.1102 &0.0847 &0.7951 &0.0837 &-0.0061 &0.0838 \cr 
50 &0.8199 &0.0689 &0.0199 &0.0717 &0.0718 &0.8033 &0.0633 &0.0041 &0.0634 &0.8200 &0.0670 &0.0200 &0.0699 &0.0693 &0.7961 &0.0598 &-0.0049 &0.0599\cr 
100 &0.8085 &0.0465 &0.0085 &0.0473 &0.0543 &0.8016 &0.0452 &0.0020 &0.0453 &0.8125 &0.0482 &0.0125 &0.0498 &0.0530 &0.8002 &0.0431 &0.0002 &0.0431\cr 
500 &0.7954 &0.0201 &-0.0046 &0.0206 &0.0258 &0.8000 &0.0199 &0.0000 &0.0198 &0.7970 &0.0184 &-0.0030 &0.0186 &0.0257 &0.7993 &0.0186 &-0.0008 &0.0186\cr 
1000 &0.7936 &0.0135 &-0.0064 &0.0149 &0.0184 &0.8002 &0.0141 &0.0002 &0.0141 &0.7961 &0.0129 &-0.0039 &0.0134 &0.0183 &0.7998 &0.0135 &-0.0003 &0.0135\cr 
    \bottomrule  
    \end{tabular}  
    \caption{Simulation results of polychoric correlation comparing the regression  method with the ML method.}\label{tab2}
    \end{threeparttable}  
\end{table} }

The MEANs of the estimators of both IRLS and ML are close to the true values of $\rho$. IRLS can be considered unbiased. The SD  of IRLS is slightly larger than that of ML, and the difference decreases as $\rho$, $s, r$ and $N$  increases, by the law of large numbers. The difference between these SDs is negligible when $\rho \geq 0.4 $, $N \geq 100$ or $s,r \geq 5$. 


The estimated standard derivation from the IRLS is smaller than the true value. However, the biases decrease with the increase of $\rho,s,r,N$. 

Because the estimator of polychoric correlation from the IRLS algorithm is determined by proportions, not the frequencies of the contingency table, increasing sample size of the data set does not reduce the speed of the estimation. In comparison, the log-likelihood function in the traditional ML methods is calculated by adding those of all data points.  Additionally, only single integrals are evaluated in the IRLS when the cumulative distribution function $\Phi$ is invoked. But the traditional ML methods need to evaluated double integrals in calculating likelihood functions.  Therefore, the IRLS is a much faster algorithm in estimating polychoric correlation comparing to the ML methods. Our simulations show that IRLS requires significantly less running time than the ML methods. Table \ref{tabtime} presents the running time of the simulation studies comparing with the ML and 2-steps method (IRLS calculate variance but the others not because they are too slow), with $\rho = 0.4, N = 500$ and 1000 replicates. The user time is the CPU time charged for the execution of user instructions of the calling process. The system time is the CPU time charged for execution by the system on behalf of the calling process. 

\setlength{\tabcolsep}{0.8mm}{
\begin{table}[H]\large
  \centering
  \fontsize{6}{5}\selectfont  
  \begin{threeparttable}  
    \begin{tabular}{ccccccccccccccccccc}
    \toprule  
    \multicolumn{1}{l}{} &\multicolumn{3}{c}{IRLS} &\multicolumn{3}{c}{ML method} &\multicolumn{3}{c}{2-step method}&\multicolumn{3}{c}{IRLS} &\multicolumn{3}{c}{ML method}&\multicolumn{3}{c}{2-step method}\cr
    \cmidrule(lr){2-4}\cmidrule(lr){5-7} \cmidrule(lr){8-10}\cmidrule(lr){11-13}\cmidrule(lr){14-16}\cmidrule(lr){17-19}
               &user  &system &elapsed       &user  &system &elapsed &user  &system &elapsed       &user  &system &elapsed  &user  &system &elapsed &user  &system &elapsed \cr 

    \cmidrule(lr){2-10} \cmidrule(lr){11-19}
    & \multicolumn{9}{c}{Polyserial correlation} &\multicolumn{9}{c}{Polychoric correlation} \cr
    \midrule
     k = 2 &0.64 &0.03 &0.67  &5.56 &0.02 &5.56 &1.11 &0.00 &1.10 &4.68 &0.05 &4.63 &31.01 &0.09 &31.16 &5.93 &0.00 &5.83\cr 
     k = 3  &0.67 &0.02 &0.69  &9.21 &0.05 &9.28 &1.02 &0.03 &1.07 &4.42 &0.05 &4.52  &113.31 &0.06 &113.35 &11.80 &0.01 &11.82\cr
     k = 5 &0.68 &0.03 &0.72  &22.59 &0.06 &22.66 &1.13 &0.03 &1.18 &4.18 &0.03 &4.16  &1756.90 &1.40 &1758.69 &31.65 &0.06 &31.61\cr
     k = 7  &0.70 &0.05 &0.73  &45.97 &0.11 &46.11 &1.05 &0.01 &1.06 &4.70 &0.00 &4.74  &2479.88 &1.74 &2482.21 &61.34 &0.04 &61.29\cr 
    \bottomrule  
    \end{tabular}  
    \caption{Running time of IRLS and ML method by Intel(R) Core(TM) i7-10700 CPU @ 2.90GHz}\label{tabtime}
    \end{threeparttable}  
\end{table}}
The functions of  the proposed algorithm is implemented in R language while the \texttt{polychor} function in the  \texttt{Polycor} package calls the \texttt{optimize} function which is written in C language for optimization. The results show that the proposed method still requires less running time than the ML and 2-steps method. In addition,  the running time of the ML and 2-steps method increases significantly, while the proposed IRLS algorithm does not slow down much,  as the number of categories getting larger.

\section{Data Analysis}
\begin{figure}[H]
  \subcaptionbox{Polychoric correlation estimated by the IRLS algorithm}{\includegraphics[width=0.48\columnwidth]{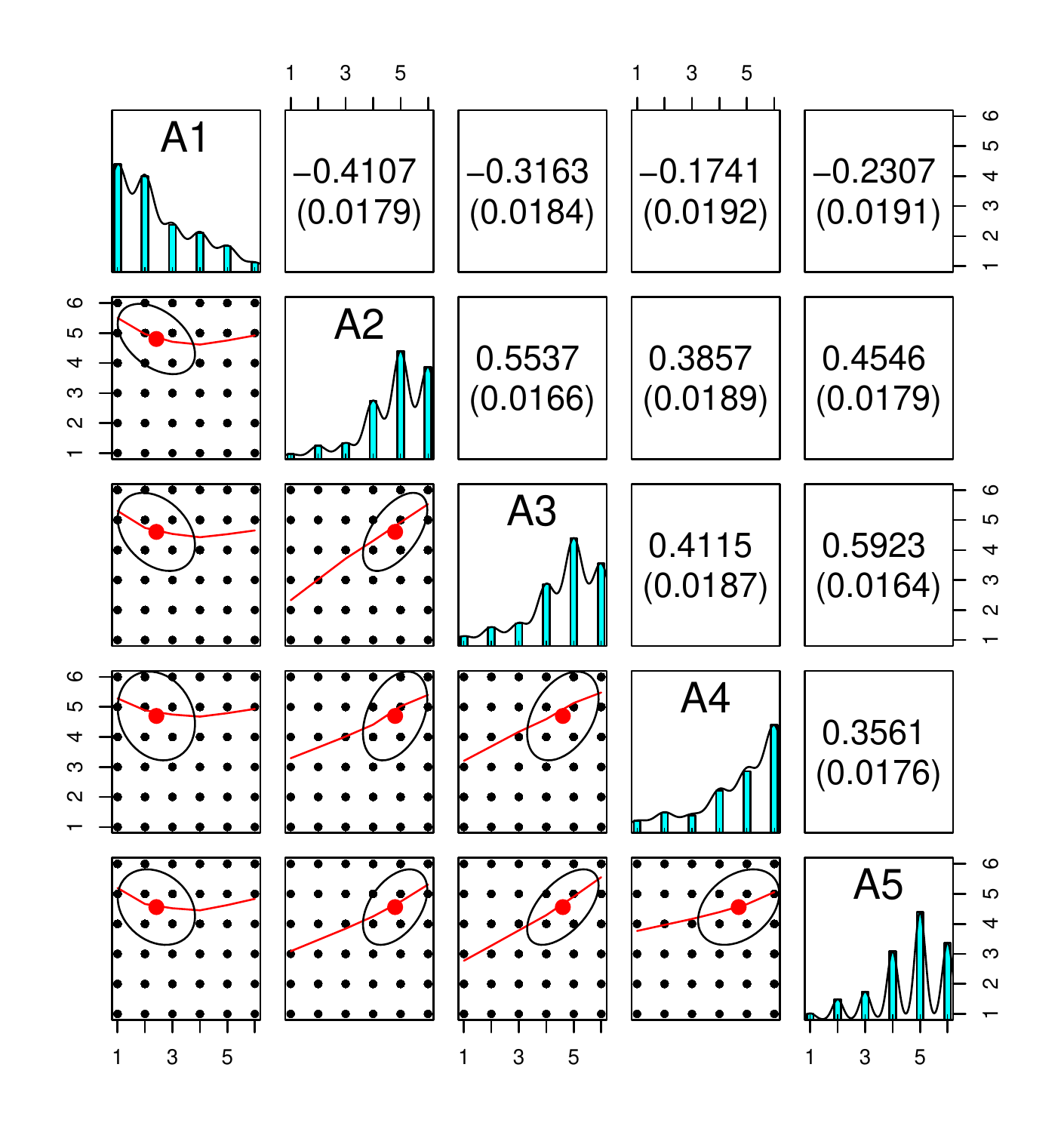}}
  \subcaptionbox{Polychoric correlation estimated by the ML method}{\includegraphics[width=0.48\columnwidth]{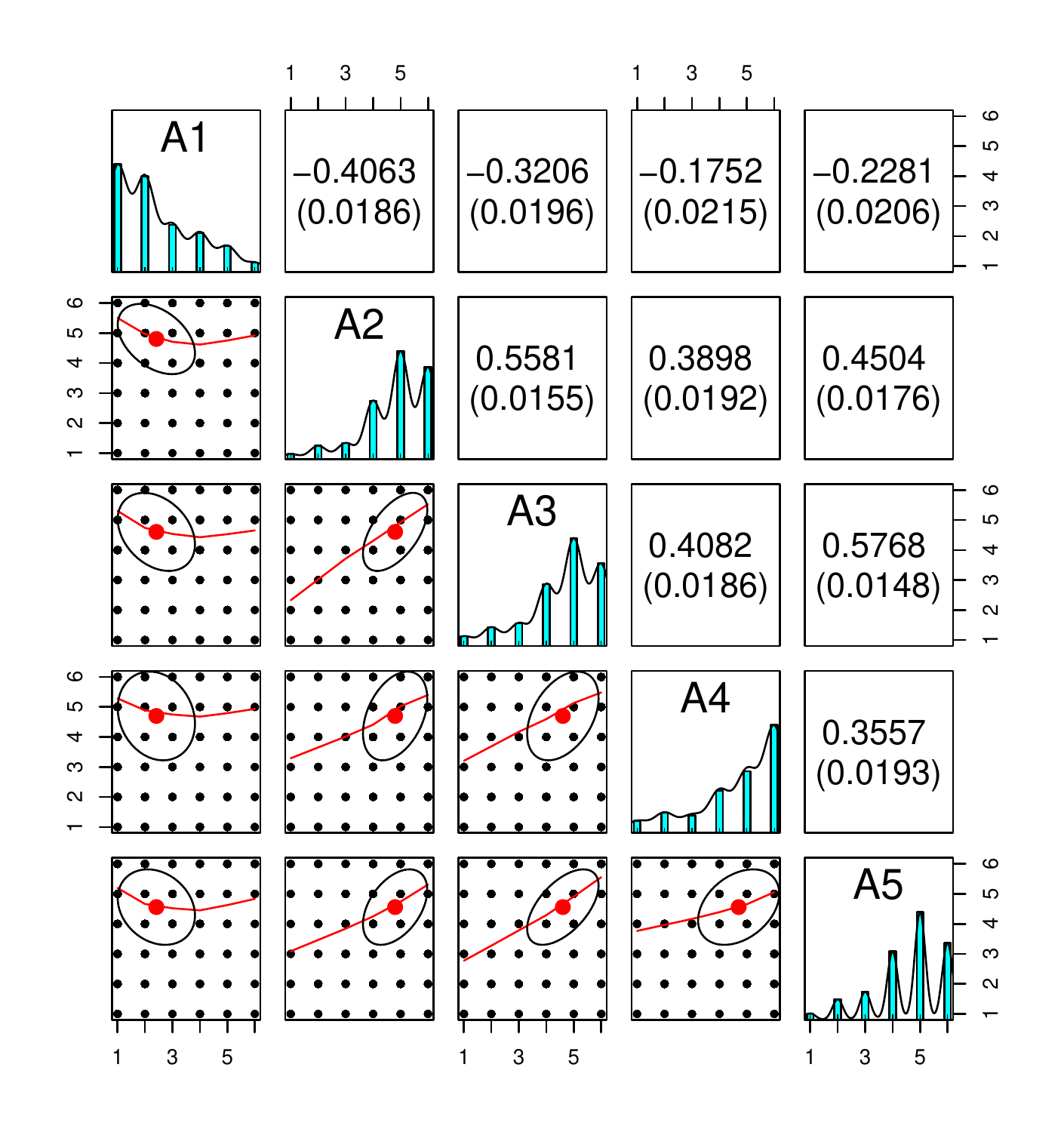}}
\caption{Polychoric correlation estimated by the IRLS algorithm and the ML method}\label{f1}
\end{figure}

In this section, two real data analyses are conducted in comparing the performance of the proposed method and the ML method, including running time recorded. The first dataset is the Big Five Inventory data provided in the \texttt{psych} package. The dataset includes 25 personality self report items taken from 2800 subjects. The item data were collected using a 6 point response scale: 1 for Very Inaccurate, 2 for Moderately Inaccurate, 3 for Slightly Inaccurate, 4 for Slightly Accurate, 5 for Moderately Accurate and 6 for Very Accurate. Three additional demographic variables (sex, education, and age) are also included. Figure \ref{f1} shows a scatter plot of matrices (SPLOM), with bivariate scatter plots below the diagonal, histograms on the diagonal, and the polychoric correlation with standard derivation above the diagonal of the first five variables. Correlation ellipses are drawn in the same graph. The red lines below the diagonal are the LOESS smoothed lines, fitting a smooth curve between two variables.

It can be seen  from Figure \ref{f1} that the estimates from the proposed method are similar to those from the ML method.

The running times are listed in Table \ref{tab7}. The user time is the CPU time spent in the execution of user instructions of the calling process. The system time is the CPU time charged for spent in the execution by the system on behalf of the calling process.  Clearly, the proposed method requires significantly less CPU time than the ML method in estimating the polychoric correlations. 
\setlength{\tabcolsep}{3.8mm}{
\begin{table}[H]
  \centering
  \fontsize{8}{6}\selectfont  
  \begin{threeparttable}  
    \begin{tabular}{cccccccccc}
    \toprule  
    \multicolumn{1}{l}{} &\multicolumn{3}{c}{IRLS} &\multicolumn{3}{c}{ML method} &\multicolumn{3}{c}{2-steps method}\cr
    \cmidrule(lr){2-4}\cmidrule(lr){5-7}\cmidrule(lr){8-10}
    N =2800           &user  &system &elapsed       &user  &system &elapsed  &user  &system &elapsed\cr 

    \cmidrule(lr){2-10}
     &1.55   &0.00   &1.55  &987.68   &1.00   &988.90 &43.60 &0.77 &44.10 \cr 

    \bottomrule  
    \end{tabular}  
    \caption{Running time of the IRLS and the ML,2-steps method}\label{tab7}
    \end{threeparttable}  
\end{table}}

The second study applies the proposed and the traditional ML methods on the data in \cite{li2019testing}, which were collected from 428 classrooms of 193 preschools from eight provinces of China using the Chinese Early Childhood Environment Rating Scale (CECERS), a newly developed quality measurement tool, to evaluate the classroom quality. The CECERS uses a 9-point scoring system, 1-3 (inadequate), 5 (least acceptable), 7 (good), and 9 (excellent), to measure the quality of Chinese early children education (ECE) programs for children aged 3 to 6. The CECERS has a total of 51 items organized in eight categories: (1) Space and Furnishings (9 items); (2) Personal Care Routines (6 items); (3) Curriculum Planning and Implementation (5 items); (4) Whole-Group Instruction (7 items); (5) Activities (9 items); (6) Language-Reasoning (4 items); (7) Guidance and Interaction (5 items); (8) Parents and Staff (6 items). 

\begin{figure}[H]
  \subcaptionbox{Polychoric correlation estimated by the IRLS  algorithm}{\includegraphics[width=0.48\columnwidth]{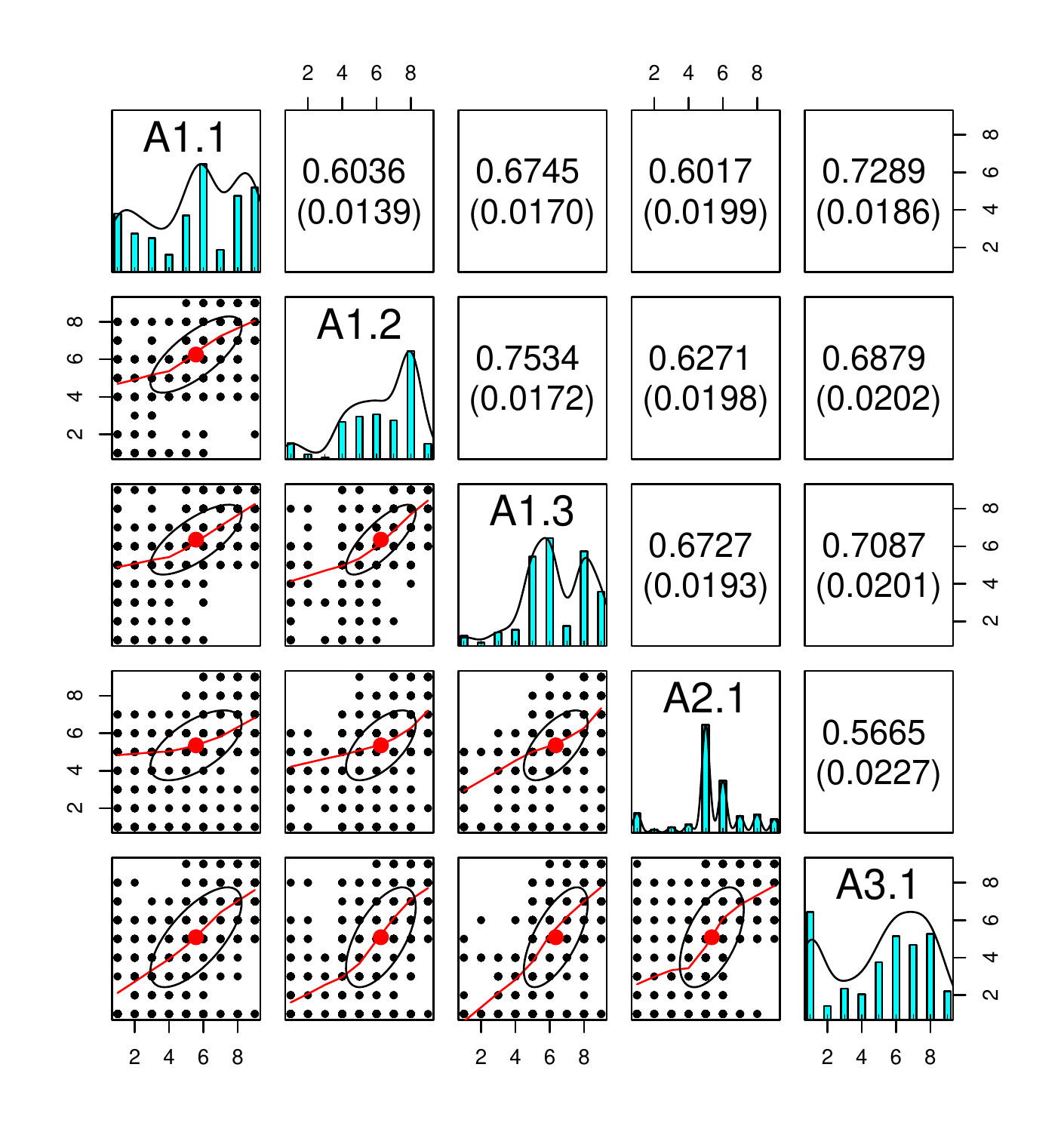}}
  \subcaptionbox{Polychoric correlation estimated by the ML method}{\includegraphics[width=0.48\columnwidth]{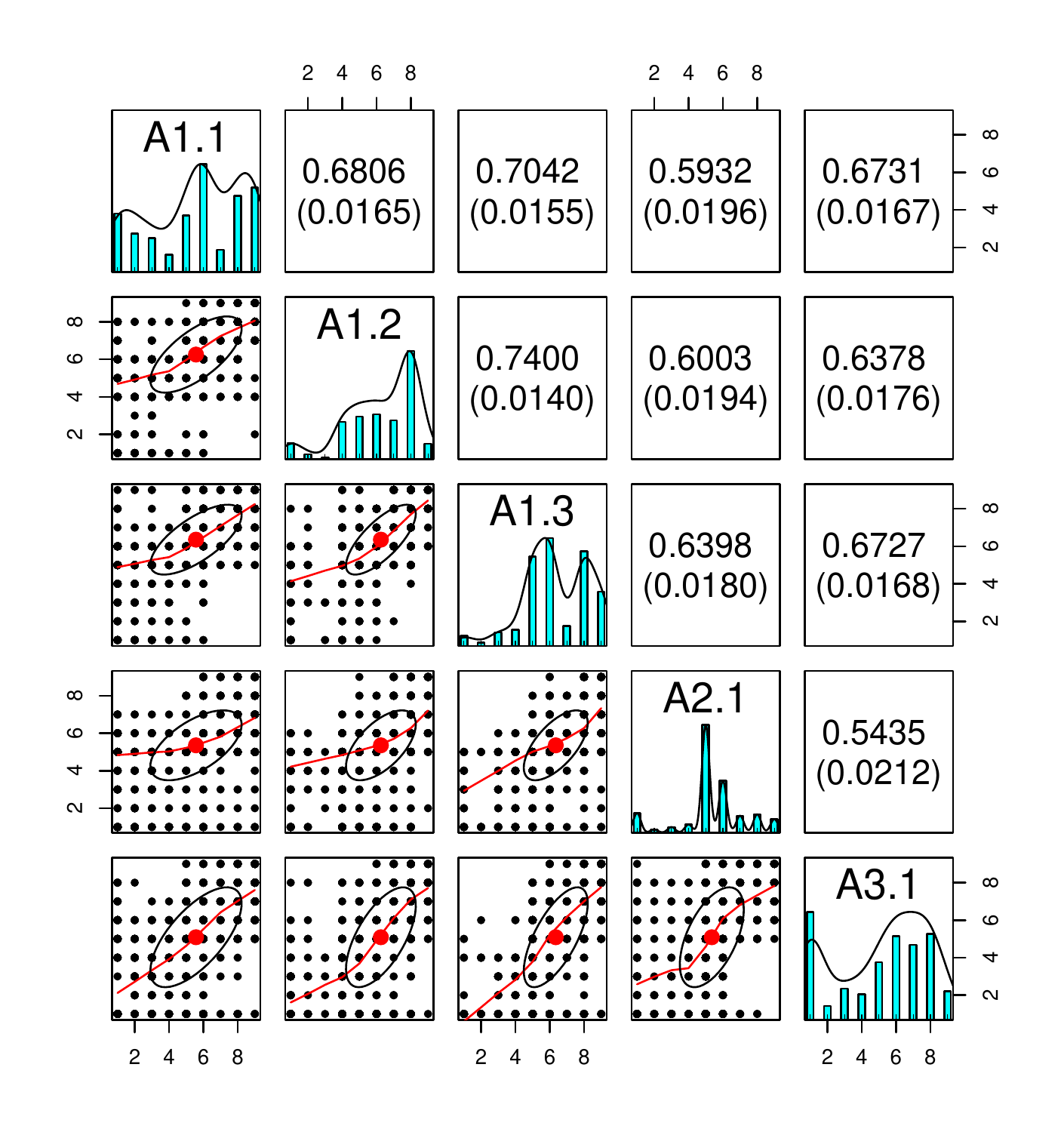}}
\caption{Polychoric correlation estimated by the IRLS  algorithm and the ML method}\label{f2}
\end{figure}

Figure \ref{f2} presents a pairwise scatter plot of matrices (SPLOM) with  the first 5 variables from the first category, Space and Furnishings, with bivariate scatter plots below the diagonal, histograms on the diagonal, and the polychoric correlation with standard derivation above the diagonal. Correlation ellipses are drawn in the same graph. The red lines below the diagonal are the LOESS smoothed lines, fitting a smooth curve between two variables. 


Table \ref{tab8} presents the running times of  two competing methods (ML is not presented because it is too slow). With moderate sample size (428) and moderate number of categories (51), the advantage of the proposed method in speed is manifested in that the running time of 2-steps method is much longer than IRLS (over 95 variables with no NAs). 

\setlength{\tabcolsep}{3.8mm}{
\begin{table}[H]
  \centering
  \fontsize{8}{6} \selectfont 
  \begin{threeparttable}  
    \begin{tabular}{cccccccccc}
    \toprule  
    \multicolumn{1}{l}{} &\multicolumn{3}{c}{IRLS} &\multicolumn{3}{c}{ML method} &\multicolumn{3}{c}{2-steps method}\cr
    \cmidrule(lr){2-4}\cmidrule(lr){5-7} \cmidrule(lr){8-10}
    N =428           &user  &system &elapsed       &user  &system &elapsed  &user  &system &elapsed\cr 

    \cmidrule(lr){2-10}
     &24.27   &0.01  &24.33   &None   &None   &None &2129.11 &10.11 &2135.99 \cr 

    \bottomrule  
    \end{tabular}  
    \caption{Running time of  the IRLS and the ML method}\label{tab8}
    \end{threeparttable}  
\end{table}}

\section{Conclusion and Discussion}
In this paper, we develop a new method to estimate polyserial and  polychoric correlation coefficients. Simulation studies and data analyses show that the proposed IRLS  algorithm can estimate polyserial and polychoric correlations consistently  and  efficiently. It  also takes much less time to compute than the traditional ML method. This prominent aspect of the new approach may help in modern research with huge datasets to analyze. The new method makes studies on big data using polyserial or polychoric correlation plausible, such as in network data and text mining.

The basic idea of the proposed method is that the correlation coefficient of two continuous variables can be obtained from the slope of the derived regression models. The regression coefficient is not necessarily estimated from the whole data. Instead, it can be calculated from  points at the conditional expected values and through which the regression line pass in theory. Hence the regression coefficient, i.e. the correlation coefficient, can be estimated with some sufficient statistics of the data. An iteratively reweighted least squares method is proposed to estimate the regression coefficient. This method  is in effect a generalized moment method based on summary statistics that always generates consistent estimators. A more general algorithm with different choices of estimating equations will be studied in the future.

The advantage of this method is obvious. It is applicable to summary data so it can be used for meta analysis that combines different studies with only summaries of data reported.  Meta analysis on the polyserial and polychoric correlation will be another future work.  

When the sample size is small, the standard errors estimated using the IRLS algorithm have more biases than the ML methods, due to the normal approximation of the delta method for calculating the variance of the regression coefficient. However, in real data applications, researchers usually just need to estimate the polyserial and polychoric correlation coefficients, there is  little interest in doing statistical inference on them. 

Another possible aspects of the proposed method left to future study is its robustness to the distributional assumption. We assumed that the  latent variables have an underlying bivariate normal distribution. Whether this is true is testable. However it is out of the scope of this paper.  
It is worthwhile to examine  what degree of departures from the normality assumption has any effects on the correlation estimation in the future.


\section{Acknowledgement}
This research was partially supported by The National Social Science Fund of China (Project No. BHA160093, titled The Design Research of Learning Environment of Primary and Secondary Schools).

\section{Appendix}
The partial derivative matrix $  \mathbf{D} $  is a $2 \times 4$ matrix,  given by 

$$ \mathbf{D} =   \mathbf{D}_1 +  \mathbf{D}_2$$ where
\begin{equation}\label{dera}
    \mathbf{D}_1 =  \left(\begin{array}{cccc} q_{12}(e_{11}-e_{12})  & q_{11}(e_{12}-e_{11})  & 0 & 0 \\
    0 & 0 & q_{22}(e_{21}-e_{22}) & q_{21}(e_{22}-e_{21}) \end{array} \right), 
\end{equation}

\begin{eqnarray*}
 q_{11} &= & \frac{P_{11}}{(P_{11}+P_{12})^2} \\
 q_{12} &=& \frac{P_{12}}{(P_{11}+P_{12})^2} \\
 q_{21} &=&  \frac{P_{21}}{(P_{21}+P_{22})^2} \\
 q_{22} &=&  \frac{P_{22}}{(P_{21}+P_{22})^2}
 \end{eqnarray*}
and
$$ \mathbf{D}_2 =  \left(\begin{array}{cccc}  p_{11}\frac{\partial{e_{11}}}{\partial{P_{11}}} + p_{12}\frac{\partial{e_{12}}}{\partial{P_{11}}} & 0 & p_{11}\frac{\partial{e_{11}}}{\partial{P_{21}}} + p_{12}\frac{\partial{e_{12}}}{\partial{P_{21}}} & 0 \\
  p_{21}\frac{\partial{e_{21}}}{\partial{P_{11}}} + p_{22}\frac{\partial{e_{22}}}{\partial{P_{11}}} & 0 & p_{21}\frac{\partial{e_{21}}}{\partial{P_{21}}}  + p_{22}\frac{\partial{e_{22}}}{\partial{P_{21}}} & 0 \end{array} \right) 
$$
\begin{eqnarray*}
 p_{11} &= & \frac{P_{11}}{P_{11}+P_{12}} \\
 p_{12} &=& \frac{P_{12}}{P_{11}+P_{12}} \\
 p_{21} &=&  \frac{P_{21}}{P_{21}+P_{22}} \\
 p_{22} &=&  \frac{P_{22}}{P_{21}+P_{22}} \\
        \frac{\partial{e_{i1}}}{\partial{P_{11}}} &= & \frac{\partial{e_{i1}}}{\partial{P_{21}}}=\frac{\phi(e_{x_i}^{b})\{e_{x_i}^{b}\Phi(e_{x_i}^{b})+\phi(e_{x_i}^{b})\}}{\Phi^2(e_{x_i}^{b})\phi(b)}\\
        \frac{\partial{e_{i2}}}{\partial{P_{11}}} & =&  \frac{\partial{e_{i2}}}{\partial{P_{21}}}=\frac{\phi(e_{x_i}^{b})[-e_{x_i}^{b}\{1-\Phi(e_{x_i}^{b})\}+\phi(e_{x_i}^{b})]}{\{1-\Phi(e_{x_i}^{b})\}^2\phi(b)}\\
        \frac{\partial{e_{ij}}}{\partial{P_{i2}}} &=& 0,(i,j=1,2)
 \end{eqnarray*}

\bibliographystyle{plainnat}
\bibliography{Tetrachoric}


\end{document}